\documentclass[useAMS,usenatbib]{mn2e}
\usepackage{graphicx,subfigure,amssymb,times}
\usepackage[normalem]{ulem}
\usepackage{soul}
\usepackage{color}
\definecolor{lightgray}{rgb}{.92,.92,.92}
\sethlcolor{lightgray}

\begin{document}

%
%
\newcommand{\wav}[1]{$\lambda#1\,\mathrm{cm}$}  
\newcommand{\wwav}[2]{$\lambda\lambda#1,#2\,\mathrm{cm}$}  
\newcommand{\wwwav}[3]{$\lambda\lambda#1,#2,#3\,\mathrm{cm}$}  
\newcommand{\wwwwav}[4]{$\lambda\lambda#1,#2,#3,#4\,\mathrm{cm}$}  
\newcommand{\Bt}{B_\mathrm{tot}} 
\newcommand{\Btpa}{B_\mathrm{tot\parallel}}
\newcommand{\Btpe}{B_\mathrm{tot\perp}}
\newcommand{\Br}{B} 
\newcommand{\BrO}{B_0}
\newcommand{\Brpa}{B_{\parallel}}
\newcommand{\Brpe}{B_{\perp}}
\newcommand{\Bur}{B_{r}} 
\newcommand{\But}{B_{\theta}}
\newcommand{\Buz}{B_{z}}
\newcommand{\bt}{b} 
\newcommand{\btO}{b_0}
\newcommand{\btpa}{b_{\parallel}}
\newcommand{\btpe}{b_{\perp}}
\newcommand{\hth}{h_\mathrm{th}}
\newcommand{\hhalo}{h_\mathrm{halo}}

\newcommand{\bra}[1]{\langle#1\rangle}  
\newcommand{\arm}{_{\mathrm{a}}} 
\newcommand{\down}{^{\mathrm{(d)}}} 
\newcommand{\el}{_{\mathrm{e}}}  
\newcommand{\up}{^{\mathrm{(u)}}}  
\newcommand{\pdif}{\partial}
\newcommand{\dif}{\,\mathrm{d}}
\newcommand{\dd}{\,\mathrm{d}}
\newcommand{\const}{\mathrm{const}}
%
%
\newcommand{\cm}{\,\mathrm{cm}}
\newcommand{\mm}{\,\mathrm{mm}}
\newcommand{\cmcube}{\,\mathrm{cm^{-3}}}
\newcommand{\dyn}{\,\mathrm{dyn}}
\renewcommand{\deg}{^{\circ}}
\newcommand{\erg}{\,\mathrm{erg}}
\newcommand{\FRM}{\,\mathrm{rad\,m^{-2}}} 
\newcommand{\g}{\,\mathrm{g}}
\newcommand{\Hz}{\,\mathrm{Hz}}
\newcommand{\GHz}{\,\mathrm{GHz}}
\newcommand{\Jy}{\,\mathrm{Jy}} 
\newcommand{\Jyb}{\,\mathrm{Jy/beam}} 
\newcommand{\cms}{\,\mathrm{cm\,s^{-2}}}
\newcommand{\kl}{\,\mathrm{k}\lambda} 
\newcommand{\kms}{\,\mathrm{km\,s^{-1}}}
\newcommand{\mJy}{\,\mathrm{mJy}} 
\newcommand{\mJyb}{\,\mathrm{mJy/beam}} 
\newcommand{\K}{\,\mathrm{K}}
\newcommand{\kpc}{\,\mathrm{kpc}}
\newcommand{\Mpc}{\,\mathrm{Mpc}}
\newcommand{\mG}{\,\mathrm{mG}} 
\newcommand{\MHz}{\, \mathrm{MHz}}
\newcommand{\Msol}{\,\mathrm{M_\sun}}
\newcommand{\n}{n_\mathrm{e}}
\newcommand{\pc}{\,\mathrm{pc}}
\newcommand{\RM}{\mathrm{RM}}
\newcommand{\DP}{\mathrm{DP}}
\newcommand{\RMi}{\mathrm{RM_i}} 
\newcommand{\RMfg}{\mathrm{RM_{fg}}} 
\newcommand{\s}{\,\mathrm{s}}
\newcommand{\uG}{\,\mu\mathrm{G}} 
\newcommand{\uJy}{\,\mu\mathrm{Jy}} 
\newcommand{\uJyb}{\,\mu\mathrm{Jy/beam}} 
\newcommand{\yr}{\,\mathrm{yr}}
%
%
\title[Magnetic fields and spiral arms in M51]
{
Magnetic fields and spiral arms in the galaxy M51}

\author[A.\ Fletcher et al.]{
        A.\ Fletcher,$^{1}$\thanks{E-mail: andrew.fletcher@ncl.ac.uk}
        R.\ Beck,$^2$
        A.\ Shukurov,$^1$
        E.\ M.\ Berkhuijsen$^2$
        and C.\ Horellou$^3$ \\
        $^1$School of Mathematics and Statistics, Newcastle University,
        Newcastle-upon-Tyne NE1 7RU, U.K. \\
    $^2$Max-Planck-Institut f\"ur Radioastronomie, Auf dem H\"ugel 69,
        53121 Bonn, Germany \\
    $^3$Onsala Space Observatory, Chalmers University of Technology, 439 92 Onsala, Sweden}

\maketitle

\begin{abstract}

We use new multi-wavelength radio observations, made with the VLA and
Effelsberg telescopes, to study the magnetic field of the nearby
galaxy M51 on scales from $200\pc$ to several $\kpc$. Interferometric
and single dish data are combined to obtain new maps at \wwav{3}{6} in
total and polarized emission, and earlier \wav{20} data are
re-reduced. We compare the spatial distribution of the radio emission
with observations of the neutral gas, derive radio spectral index and
Faraday depolarization maps, and model the large-scale variation in
Faraday rotation in order to deduce the structure of the regular
magnetic field. We find that the \wav{20} emission from the disc is
severely depolarized and that a dominating fraction of the observed
polarized emission at \wav{6} must be due to anisotropic small-scale
magnetic fields. Taking this into account, we derive two components
for the regular magnetic field in this galaxy: the disc is dominated
by a combination of azimuthal modes, $m=0+2$, but in the halo only an
$m=1$ mode is required to fit the observations. We disuss how the
observed arm-interarm contrast in radio intensities can be reconciled
with evidence for strong gas compression in the spiral shocks. In the
inner spiral arms, the strong arm--interarm contrasts in total and
polarized radio emission are roughly consistent with expectations from
shock compression of the regular and turbulent components of the
magnetic field. However, the average arm--interam contrast,
representative of the radii $r>2\kpc$ where the spiral arms are
broader, is not compatible with straightforward compression: lower
arm--interarm contrasts than expected may be due to resolution effects
and \emph{decompression} of the magnetic field as it leaves the arms.
We suggest a simple method to estimate the turbulent scale in the
magneto-ionic medium from the dependence of the standard deviation of
the observed Faraday rotation measure on resolution. We thus obtain an
estimate of $50\pc$ for the size of the turbulent eddies.

\end{abstract}

\section{Introduction}
\label{sec:intro}

The Whirlpool galaxy, M51 or NGC\,5194, is one of the classical
grand-design spiral galaxies. The two spiral arms of M51 can be
traced through more than 360\degr\ in azimuthal angle in numerous
wavebands. M51 is probably perturbed by a recent encounter with its
companion galaxy NGC\,5195. Such interactions usually result in
enhanced star formation, either localized or global, as tidal forces
and density waves compress the interstellar medium. In M51 this may have
resulted in two systems of density waves \citep{Elmegreen89}.

M51 was the first external galaxy where polarized radio emission was detected
\citep{Mathewson72} and one of the few external galaxies where optical polarization has been studied \citep{Scarrott:1987}. \citet{Neininger:1992} and \citet{Horellou:1992} found that
the lines of the regular magnetic field in M51 have a spiral shape, but their
resolution was too low to determine how well the field is aligned with the
optical spiral arms. \citet{Horellou:1992} realized that, at wavelengths of
$\lambda\ge18$~cm, only polarized emission from a foreground layer reaches the
observer because of Faraday depolarization, and the Faraday rotation measures
obtained at the longer wavelengths are much smaller than those at shorter
wavelengths. \citet{Heald:2009} observed a fractional polarization at \wav{22} of around $5\%$
in the optical disc, increasing to $30\%$ at large radii, and
Faraday rotation measures of a similar magnitude to those found by
\citet{Horellou:1992}.  \citet{Berkhuijsen:1997} analyzed the global magnetic
field of M51 using radio data at four frequencies and found that the average
orientation of the fitted magnetic field is similar to the average pitch angle
of the optical spiral arms measured by \citet{Howard90}.
\citet{Berkhuijsen:1997} represented the magnetic field in the disc of M51 as a
superposition of periodic azimuthal modes, with about equal contribution from
the axisymmetric $m=0$ and the bisymmetric $m=1$ ones. Their fit contains a
magnetic field reversal at about 5~kpc radius which extends over a few kpc in azimuth.
Furthermore, \citet{Berkhuijsen:1997} found evidence for an axisymmetric
magnetic field in the halo of M51 (visible at \wav{18} and \wav{20}) with a
reversed direction (inwards) with respect to the axisymmetric mode of the disc
field (outwards).

The results of \citet{Berkhuijsen:1997} indicate that the magnetic
field of M51 is strong and partly regular, with some interesting
properties. However, some of their results may be affected by the
low angular resolution which was limited by the Effelsberg
single-dish data at \wav{2.8}. Furthermore, no correction for
missing large-scale structure could be applied to the VLA \wav{6}
polarization data because no single-dish map was available at that
time.

Here we present a refined analysis based on new data with higher
resolution and better sensitivity. Our new surveys of M51 in total
and linearly polarized \wav{3} and \wav{6} radio continuum
emission combines the resolution of the VLA with the sensitivity of
the Effelsberg single-dish telescope. The new maps are of comparable
resolution to the maps of the CO(1--0) emission \citep{Helfer03} and
mid-infrared dust emission \citep{Sauvage96,Regan06} and are of
unprecedented sensitivity. The shape of the spiral radio arms and
the comparison to the arms seen with different tracers were
discussed in a separate paper \citep{Patrikeev06}.

The spiral shocks in M51 are strong and regular
\citep{Aalto99} and offer the possibility to compare arm--interarm
contrasts of gas and the magnetic field. We analyse our new
data to try to separate the contribution to the observed polarized emission from
regular (or mean) magnetic fields and anisotropic random magnetic fields
produced by compression and/or shear in the spiral arms.

The new high-resolution polarization maps allow us for the first time to
investigate in detail the interaction between the magnetic fields and the shock
fronts. Results from the barred galaxies NGC~1097 and NGC~1365 showed that the
small-scale and the large-scale magnetic field components behave differently,
i.e. the small-scale field is compressed significantly in the bar's shock, while
the large-scale field is hardly compressed \citep{Beck:2005}. This was
interpreted as a strong indication that the large-scale field is coupled to the
warm, diffuse gas which is only weakly compressed. We investigate whether a
similar decoupling of the regular magnetic field from the dense gas clouds is
suggested by the observed arm--interarm contrast in M51.

The basic parameters we adopt for M51 are: centre's right ascension
$\mathrm 13^\mathrm{h}29^\mathrm{m}52\fs709$ and declination (J2000)
$+47\degr 11\arcmin 42\farcs 59$ \citep{Ford85}; distance 7.6 Mpc,
\citep{Ciardullo:2002}, thus $1"\approx 37\pc$; position angle of major axis $-10\degr$
($0\degr$ is North) and inclination $-20\degr$ ($0\degr$ is face-on)
\citep{Tully74}. The inclination is measured from the galaxy's
rotation axis to the line-of-sight, viewed from the northern end
of the major axis, and its sign is important for the geometry of the
model discussed in Sect.~\ref{sec:B}.

\section{Observations and data reduction}
\label{sec:obs}

\begin{figure*}
\centering
\includegraphics[width=0.48\textwidth]{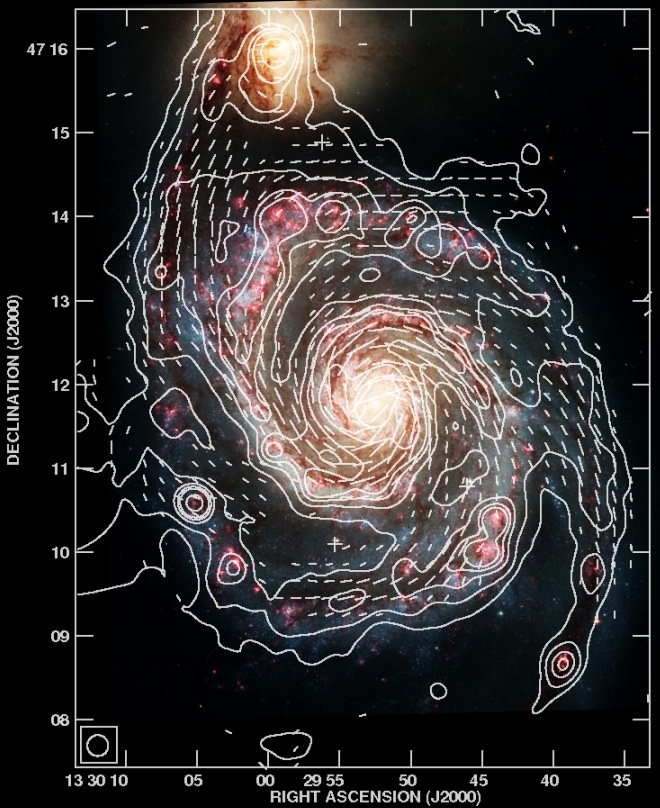}
\hfill
\includegraphics[width=0.48\textwidth]{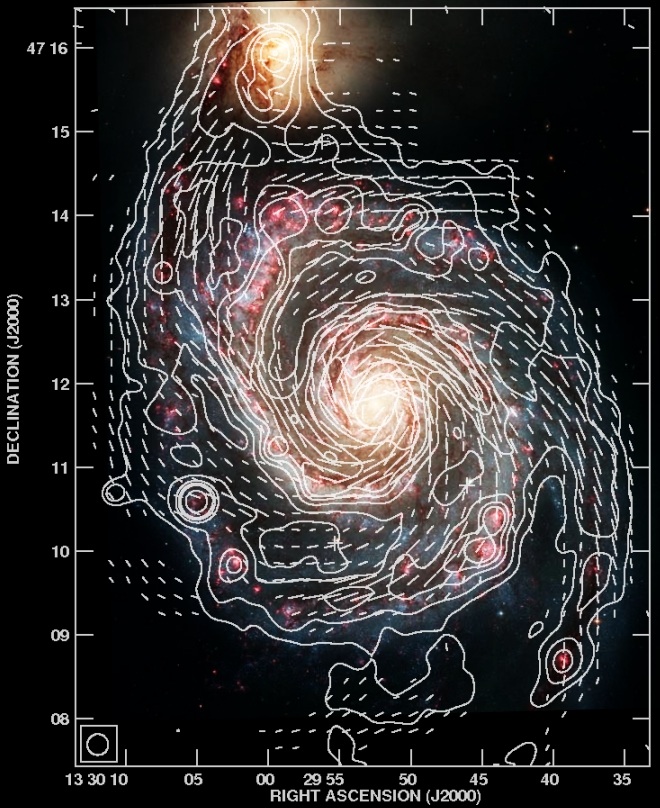}
\caption{\label{fig:i15}
\textbf{(a)} \wav{3} and \textbf{(b)} \wav{6} radio
emission at $15\arcsec$ resolution from VLA and Effelsberg
observations, overlaid on a Hubble Space Telescope optical
image (image credit: NASA, ESA,
S.~Beckwith (STScI), and The Hubble Heritage Team (STScI/AURA)). Total
intensity contours in both maps are at $6, 12, 24,
36, 48, 96, 192$ times the noise levels of $20\uJyb$ at \wav{3} and
$30\uJyb$ at \wav{6}. (Note that the roughly horizontal contours at
the left edge of panel (a) are artefacts arising from mosaicing the
two VLA pointings.) Also shown are the $B$-vectors of polarized
emission: the plane of polarization of the observed electric field
rotated by 90\degr, not corrected for Faraday rotation, with a
length proportional to the polarized intensity ($PI$) and only
plotted where $PI\ge 3\sigma_{PI}$.}
\end{figure*}

\subsection{VLA observations}

M51 was observed in October 2001 at \wav{3.5} with the
VLA\footnote{The VLA is operated by the NRAO. The NRAO is a facility
of the National Science Foundation operated under cooperative
agreement by Associated Universities, Inc.} in the compact D
configuration and in August 2001 at \wav{6.2} in the C configuration.
Two pointings of the array, at the northern and southern parts of the
galaxy, were required to obtain complete coverage of M51 at \wav{3.5}.
The data were edited, calibrated and imaged using standard AIPS
procedures and VLA calibration sources. After initial examination of
the data and the flagging of bad visibilities, self-calibration was
used. After correction for the pattern of the primary beam the two
pointings at \wav{3.5} were mosaiced using the AIPS task
\texttt{LTESS}. The calibrated \wav{6.2} C-array uv-data were combined
with existing \wav{6.2} D-array data \citep{Neininger:1996}.

A new \wav{20} map in total power and polarization is
presented in Fig.~\ref{fig:20}, based on the C-array data of
\citet{Neininger:1996} combined with D-array data of
\citet{Horellou:1992}. All of the observations were re-reduced for
this work, the two data sets were combined  and then smoothed to a
resolution of $15\arcsec$ to improve the signal-to-noise ratio. The
resulting maps at \wav{20} contain information on all scales down to
the beamsize as the primary beam of the VLA at \wav{20} in the
D-array is $\sim 30\arcmin$, about twice the size of M51 on the sky.

Different weighting schemes were used in the final imaging of the data to
produce maps with either high resolution (natural weighting) or high
signal-to-noise (uniform weighting) and for a compromise between the two
extremes (robust weighting, obtained by setting the parameter \texttt{ROBUST=0}
in the AIPS task \texttt{IMAGR}). Maps of the Stokes parameters $I$, $Q$ and $U$
were produced in each case. The optimum number of iterations of the CLEAN
algorithm used to produce the images was determined individually for each map.
Following slight smoothing, the $Q$ and $U$ maps were combined to give the
polarized intensity $PI=\sqrt{Q^2 + U^2}$, using a first-order correction for the
positive bias \citep{Wardle74}.

\begin{table}
  \begin{center}
  \caption{\label{tab:obs}
The merged VLA
and Effelsberg maps discussed in this paper, their half-power beam
widths, imaging weighting schemes and their r.m.s. noises in total
and polarized emission, $\sigma_{I}$ and $\sigma_{PI}$, respectively}
  \begin{tabular}{ccccc}
  \hline
  $\lambda$ & HPBW & Weighting & $\sigma_{I}$ & $\sigma_{PI}$\\
  (cm) & & & ($\uJyb$) & ($\uJyb$) \\
    \hline
     \phantom{2}3 &  \phantom{1}8\arcsec\  & robust   & $12$ & $10$ \\
     \phantom{2}3 & 15\arcsec\ & natural  & $20$ & $ \phantom{1}8$ \\
     \phantom{2}6 &  \phantom{1}4\arcsec\  & uniform  & $15$ & $10$ \\
     \phantom{2}6 &  \phantom{1}8\arcsec\  & robust   & $25$ & $10$ \\
     \phantom{2}6 & 15\arcsec\ & natural  & $30$ & $10$ \\
               20 & 15\arcsec\ & natural & $20$ & $13$ \\
   \hline
  \end{tabular}
  \end{center}
\end{table}

\begin{figure*}
\centering
\includegraphics[width=0.49\textwidth]{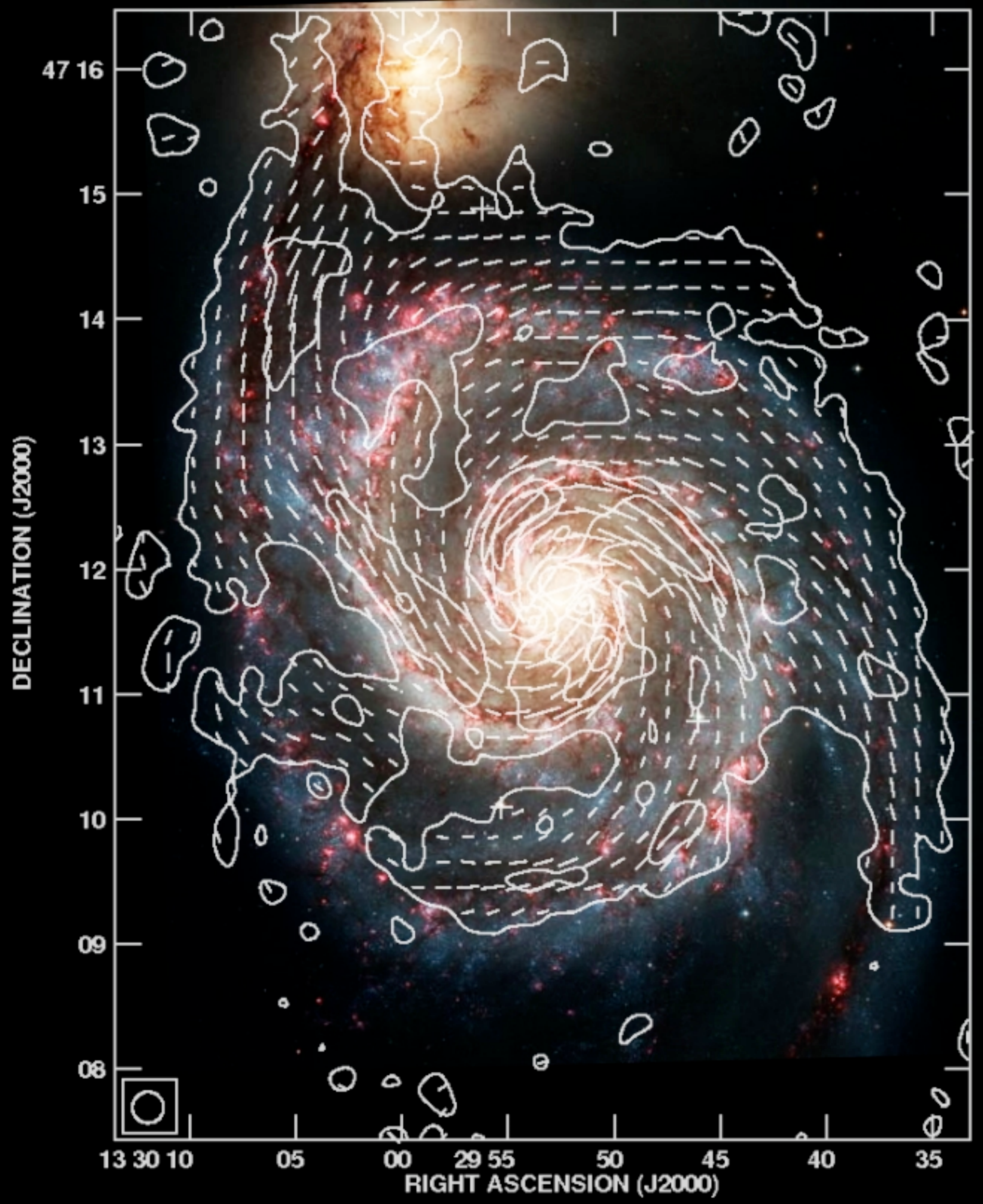}
\hfill
\includegraphics[width=0.49\textwidth]{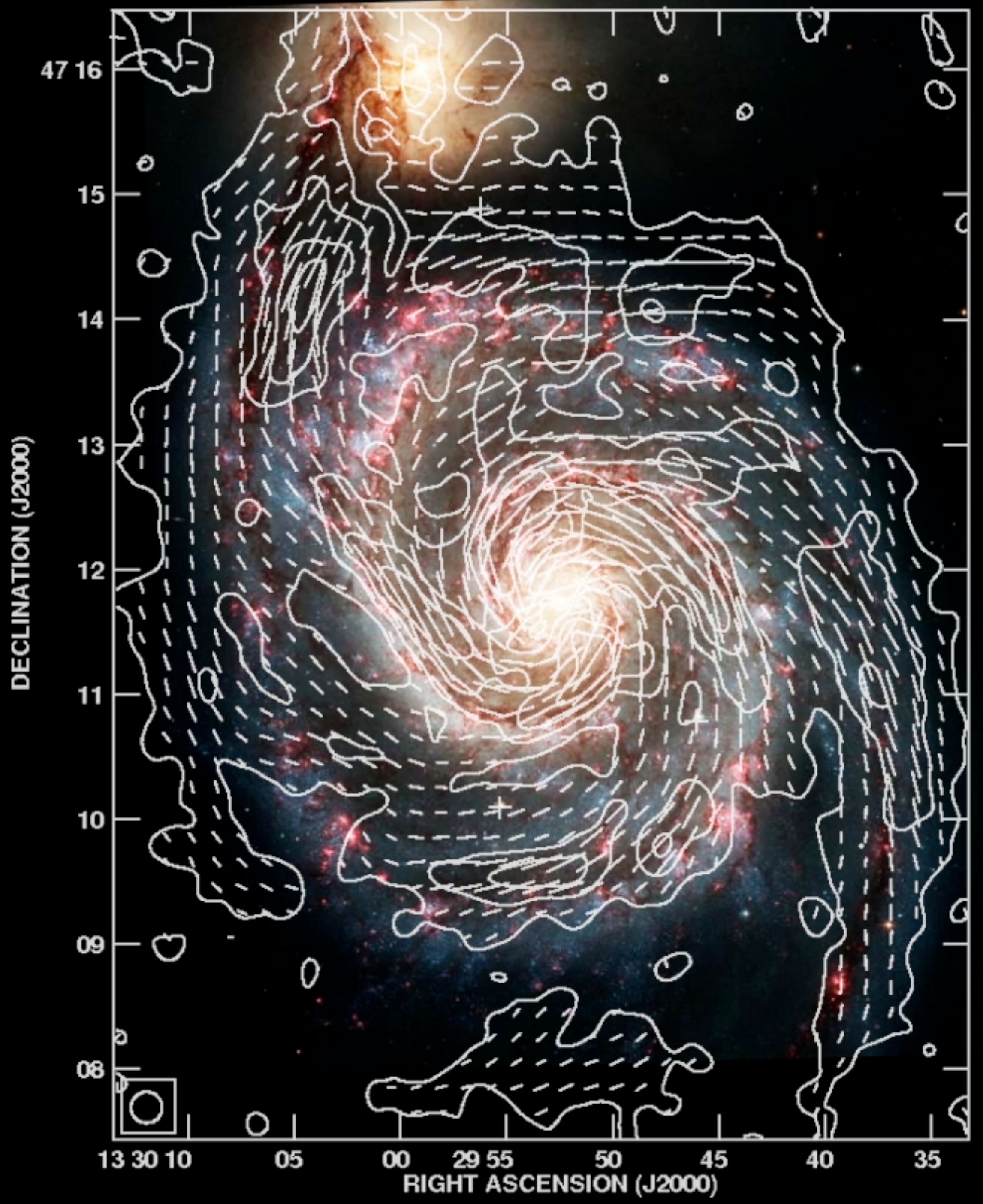}
\caption{\label{fig:pi15}
\textbf{(a)} \wav{3} and \textbf{(b)}
\wav{6} (right) polarized radio emission at $15\arcsec$ resolution
from VLA and Effelsberg observations, overlaid on the same optical
image as in Fig.~\protect\ref{fig:i15}. Polarized intensity contours in both
maps are at $3,9,15,21$ times the noise level of $8\uJyb$ at \wav{3} and $10\uJyb$ at \wav{6}. Also
shown are the $B$-vectors of polarized emission: the position angle
of the polarized electric field rotated by 90\degr, not corrected
for Faraday rotation, with the length proportional to the polarized intensity
$PI$ and only plotted where $P \ge 3\sigma_{PI}$.}
\end{figure*}

\begin{figure}
\centering
\includegraphics[width=0.47\textwidth]{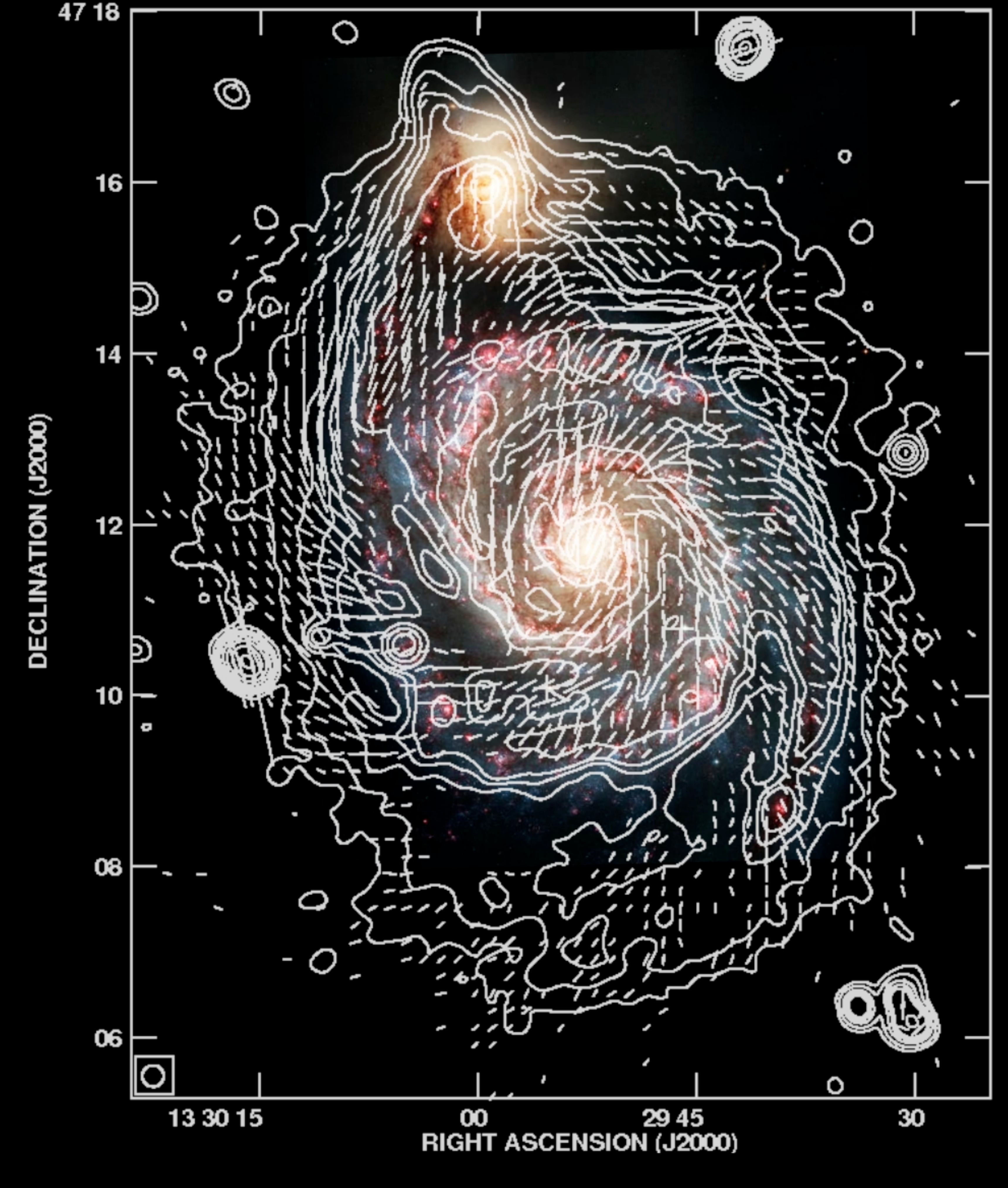}
\caption{\label{fig:20}
Contours of \wav{20}
total radio emission at $15\arcsec$ resolution, overlaid on the same
optical image as in Fig.~\protect\ref{fig:i15}. Total intensity
contours are at 6, 12, 24, 36, 48, 96, 192 times the noise level of
$20\uJyb$. Also shown are the $B$-vectors of polarized emission: the
plane of polarization of the observed electric field rotated by
90\degr, not corrected for Faraday rotation, with a length
proportional to the polarized intensity
$PI$ and only plotted where $PI\ge 3\sigma_{PI}$.}
\end{figure}

\subsection{Effelsberg observations and merging}

In order to correct the VLA maps at \wav{3} for missing extended emission we
made observations of M51 in total intensity and polarization with
the 100-m Effelsberg telescope \footnote{The Effelsberg 100-m
telescope is operated by the Max-Planck-Institut f\"ur
Radioastronomie on behalf of the Max-Planck-Gesellschaft.} in
December 2001 and April 2002 using the sensitive \wav{3.6}\ (1.1~GHz
bandwidth) receiver. We obtained 44 maps of a 12$\arcmin$ by
12$\arcmin$ field around M51, scanned in orthogonal directions. Each
map in $I$, $Q$ and $U$ was edited and baseline corrected
individually, then all maps in each Stokes parameter were combined
using a basket weaving method \citep{Emerson88}. The r.m.s. noise in
the final maps, after slight smoothing to $90\arcsec$, is $200\uJyb$
in total intensity and $20\uJyb$ in polarized intensity.

At \wav{6.2}, ten
maps of a 41$\arcmin$ by 34$\arcmin$ field were observed in November
2003 with the 4.85~GHz (500~MHz bandwidth) dual-horn receiver. The
combined data resulted in a new $180\arcsec$
resolution image with r.m.s. noise of $250\uJyb$ in
total intensity and $25\uJyb$ in polarization.

Maps from the VLA and Effelsberg were combined using the AIPS task
\texttt{IMERG}. A useful description of the principles of merging single dish
and interferometric data is given by \citet{Stanimirovic02}. The range of
overlap in the uv plane between the two images (parameter \texttt{uvrange} in
\texttt{IMERG}) was estimated as follows: we assumed an effective Effelsberg
diameter of about 60~m to estimate the maximum extent of the single dish in the
uv space ($1.7\kl$ at \wav{3.6} and $1.0\kl$ at \wav{6.2}) and used the minimum
separation of the VLA antennas in the D-array configuration, $35$~m, to
calculate the minimum coverage of the interferometer in the uv space ($1.0\kl$
at \wav{3.6} and $0.6\kl$ at \wav{6.2}). We then varied these parameters in
order to find the optimum overlap in the uv space by comparing the integrated
total flux in the merged maps with that of the single dish maps; the optimal
uv-ranges for merging were found to be $1.0\rightarrow1.6\kl$ at \wav{3.6} and
$0.5\rightarrow0.7\kl$ at \wav{6.2}. Merging of the $Q$ and $U$ maps was carried
out using the same optimum uv range as found for $I$.

The fraction of total emission (Stokes $I$) present in the VLA maps -- those
produced using natural weighting, and hence with the highest signal-to-noise
ratio -- is about 30\% at \wav{3} and close to 50\% at \wav{6} compared to the
merged maps. Small-scale fluctuations in $Q$ and $U$ due to variations in the
magnetic field orientation and Faraday rotation in M51 mean that the polarized
emission is less severely affected by missing large-scales (alternatively, the
single dish detects the large-scale emission missed by an interferometer
\emph{but} simultaneously suffers from stronger wavelength-independent beam
depolarization). At \wav{3} the VLA map contains about 75\% of the polarized
emission present in the merged map, with about 85\% present at \wav{6}.

Following the merging, the maps in $I$, $Q$ and $U$ were convolved with a
Gaussian beam to give a slightly coarser resolution and higher signal-to-noise
ratio. The maps that are discussed in this paper are listed in
Table~\ref{tab:obs} along with the r.m.s. noises in total and polarized
intensity.

\section{The M51 maps}
\subsection{The spiral arms of M51 as seen in radio emission}
\label{sec:desc}

Figure~\ref{fig:i15} shows the total radio continuum emission and the
$B$-vectors of polarized emission (the observed plane of linear
polarization rotated by 90\degr) at \wav{3} and \wav{6}, overlaid on a
Hubble Space Telescope optical image. At the assumed distance of
7.6\,Mpc, the $15\arcsec$ resolution corresponds to $560\pc$ and
$590\pc$ along the major and minor axes, respectively. The
distribution of polarized emission at \wwav{3}{6} is shown in
Fig.~\ref{fig:pi15} at $15\arcsec$ resolution. The extensive \wav{20}
total emission disc is shown in Fig.~\ref{fig:20} also at $15\arcsec$
resolution.

The total emission at \wav{3} and \wav{6} in Fig.~\ref{fig:i15} shows
a close correspondence with the optical spiral arms whereas the
\wav{20} total emission (Fig.~\ref{fig:20}) and \wav{3} and \wav{6}
polarized emission (Fig.~\ref{fig:pi15}) are spread more evenly across
the galactic disc. Compact, bright peaks of total emission coincide
with complexes of H\,{\sc ii} regions in the spiral arms, as expected
if thermal bremsstrahlung is a significant component of the radio
signal at these peaks. The flatter spectral index ($\alpha\lesssim
0.6$ where $I\propto \nu^{-\alpha}$, Fig.~\ref{fig:spx}) and the
absence of the corresponding peaks in polarized radio emission
(Fig.~\ref{fig:pi15}) suggest that a significant proportion of the
centimeter wavelength radio emission in these peaks is thermal. The
extended \wav{20} total emission and \wav{3} and \wav{6} polarized
emission accurately trace the synchrotron component of the radio
continuum.

Ridges of enhanced polarized emission are prominent in the inner
galaxy; some are located on the optical spiral arms but others are
located in-between the arms. A detailed analysis of the location and
pitch angles of the spiral arms traced by different observations and a
comparison of the pitch angles with the orientation of the regular
magnetic field is given by \citet{Patrikeev06}. They find systematic
shifts between the spiral ridges seen in polarized and total radio
emission, integrated CO line emission and infrared emission, which are
consistent with the following sequence in a density wave picture:
firstly, shock compresses gas and magnetic fields (traced by polarized
radio emission), then molecules are formed (traced by CO) and finally
thermal emission is generated (traced by infrared).
\citet{Patrikeev06} also show that while the pitch angle of the
regular magnetic field is fairly close to that of the gaseous spiral
arms at the location of the arms, the magnetic field pitch angle
changes from the above by around $\pm15\degr$ in interarm regions.

\subsection{The connection between polarized radio emission and gaseous
spiral arms} \label{sec:conn2}

\begin{figure}
\begin{center}
\includegraphics[width=0.45\textwidth]{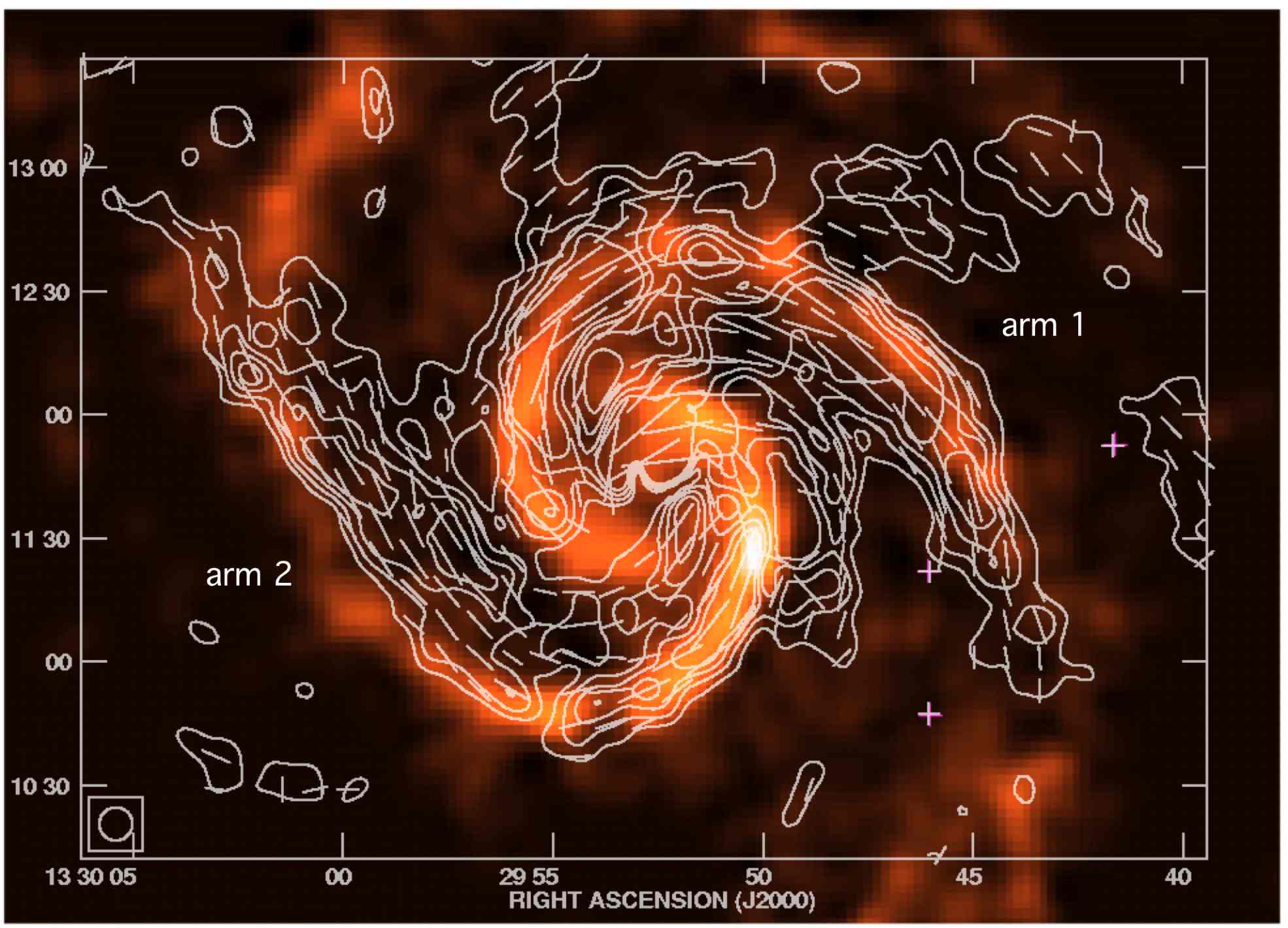}
\caption{
Contours of the \wav{6} polarized radio emission (VLA and Effelsberg combined)
in the central $\sim 3\arcmin \times 4\arcmin$ of M51 at $8\arcsec$
resolution, along with Faraday-rotation corrected B-vectors,  overlaid on the map of integrated CO(1--0)
line emission of \citet{Helfer03}. Contours are at $3,5,7,9$
times the noise level of $10\uJyb$.}
\label{fig:inner:co}
\end{center}
\end{figure}

In Fig.~\ref{fig:inner:co}, the \wav{6} polarized emission and the orientation
of the regular magnetic field in the central $8\kpc$ of M51 are overlaid onto an
image of the spiral arms as traced by the CO(1--0) integrated line emission.
Part of the polarized emission appears to be concentrated in elongated arm-like
structures that sometimes coincide with the gas spiral.

The correspondence between polarized and CO arms is good along most of the
northern arm in Fig.~\ref{fig:inner:co} (called Arm~1 in the rest of the paper)
and in the inner part of the southern arm (called Arm~2). Arm~1 continues
towards the south where the polarized emission longer coincides with the CO and
optical arms, but becomes broader further out (Fig.~\ref{fig:pi15}).

Moving along Arm~2, the excellent
overlap of the radio polarization and CO in the inner arm ends
abruptly further in the south (Fig.~\ref{fig:inner:co}), beyond
which the distribution of polarized emission is very broad and
located on the galactic centre side of the gaseous spiral arm: the peak of
this interarm polarized emission corresponds very closely with the 
pronounced peaks in radial velocity, $v_r\simeq 100\kms$, derived 
by \citet{Shetty:07} (shown in their Fig.~14). This may be an 
indication that strong shear in the interarm gas flow is producing 
this magnetic feature. At
about 6~kpc radius, the polarization Arm~2 \emph{crosses} the CO and
optical
arms in the east, followed in the northeast by a bending away from
the optical arm towards the companion galaxy NGC\,5195
(Fig.~\ref{fig:pi15}). The northern part
of Arm~2 is weakly polarized and rather irregular in total and CO
emission. The whole space between the northern
Arm~2 and the companion galaxy is filled with highly polarized radio
emission (typically 15\% at \wav{6}). Arm~2 becomes well organized
again at larger radii (located at the western edge of
Fig.~\ref{fig:pi15}), where the total radio, polarized radio and CO
emission perfectly coincide.

West of the central region, between Arms~1 and 2 in
Fig.~\ref{fig:inner:co}, another polarization feature emerges which
appears similar to the magnetic arms observed e.g. in NGC\,6946
\citep{Beck96}. However, in contrast to NGC\,6946, Faraday
rotation is not enhanced in the interarm feature of M51 (see
Fig.~\ref{fig:rm:15}). Some peaks of polarized emission between
Arms~1 and 2 in the south and southeast (see
a low-resolution image
of Fig.~\ref{fig:pi15}) and may indicate the outer extension of this
magnetic arm. Inside of the inner corotation radius, located at
4.8~kpc \citep{Elmegreen89}, this phenomenon can be explained by enhanced dynamo
action in the interarm regions \citep{Moss98, Shukurov98, Rohde99}.

\subsection{Polarized radio emission from the inner arms and central region}
\label{sec:inner}

\begin{figure*}
\begin{center}
\includegraphics[width=0.49\textwidth]{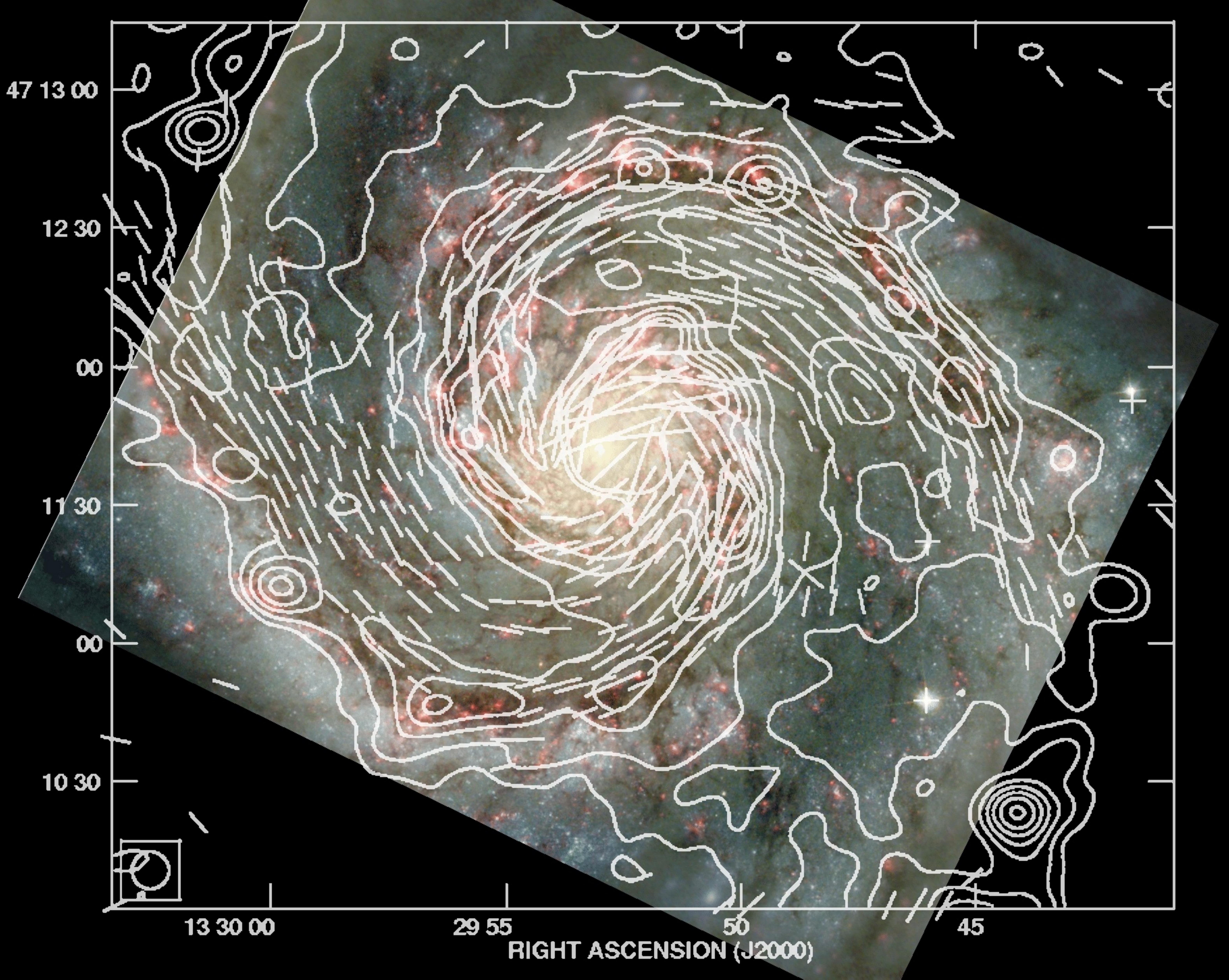}
\hfill
\includegraphics[width=0.49\textwidth]{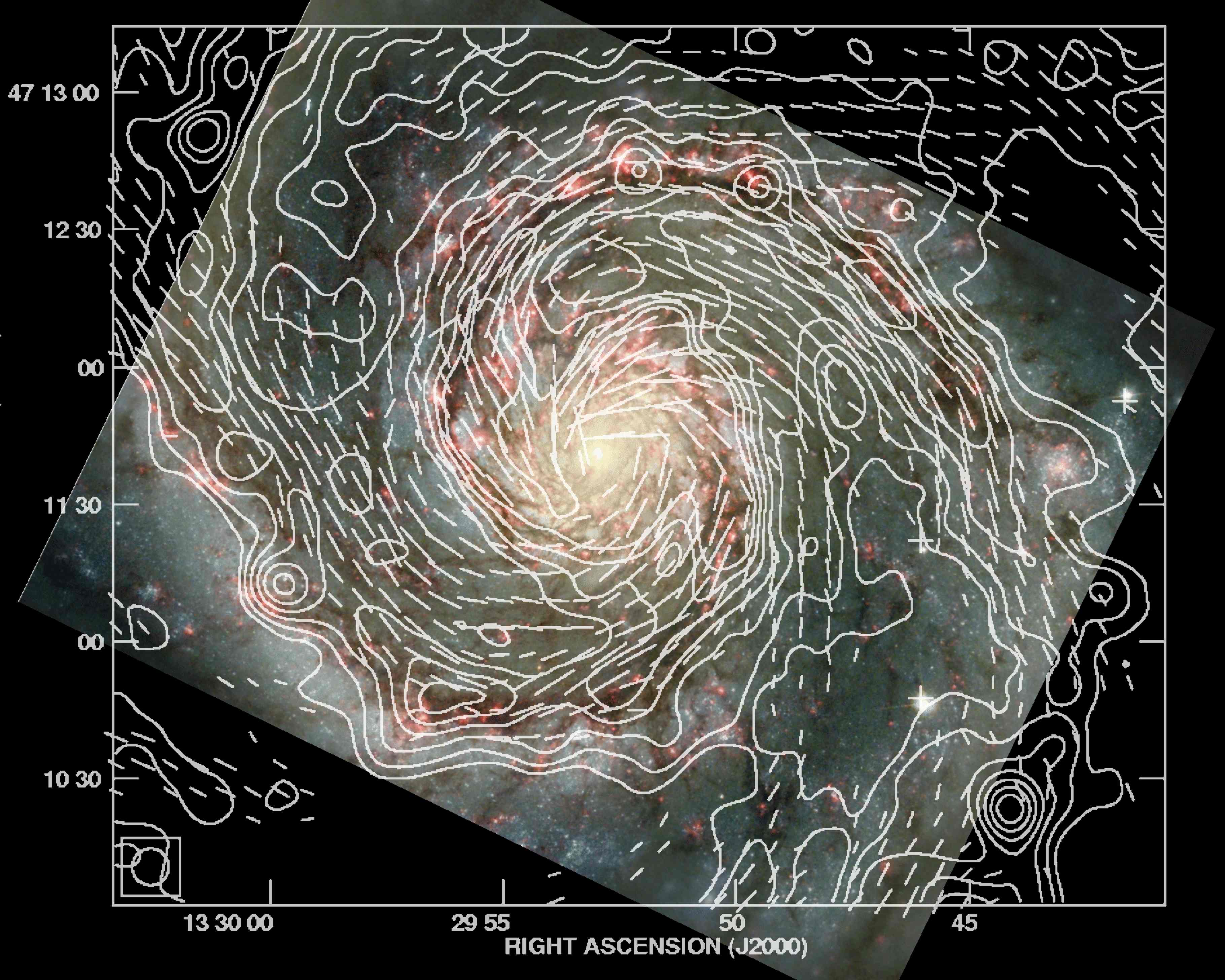}
\caption{ \label{fig:inner:8}
\textbf{(a)} \wav{3} and \textbf{(b)} \wav{6} total radio emission and 
$B$-vectors (VLA and Effelsberg combined) in the central
$3\arcmin \times 4\arcmin$ of M51 at $8\arcsec$ resolution overlaid on a Hubble
Space Telescope image ({http://heritage.stsci.edu/2001/10}, image credit: NASA
and The Hubble Heritage Team (STScI/AURA)). Contours are at $6, 12, 24, 36, 48,
96, 192$ times the noise level. 
$B$-vectors, not corrected for Faraday rotation, are plotted
where $PI\ge 3\sigma_{PI}$.}
\end{center}
\end{figure*}

In the CO and H$\alpha$ line emissions (Fig.~\ref{fig:inner:co} and the red
regions in Fig.~\ref{fig:inner:8}), the spiral arms continue towards the galaxy
center. The high-resolution CO map by \citet{Aalto99} shows that the arms are
sharpest and brightest between about $25\arcsec$ and $50\arcsec$
distance from the center. The arms become significantly broader
and less pronounced inside a radius of about 0.8~kpc: this is inside the inner
Lindblad resonance of the inner density-wave system at $r\approx 1.3\kpc$
identified by \citet{Elmegreen89}.

\begin{figure}
\begin{center}
\includegraphics[width=0.47\textwidth]{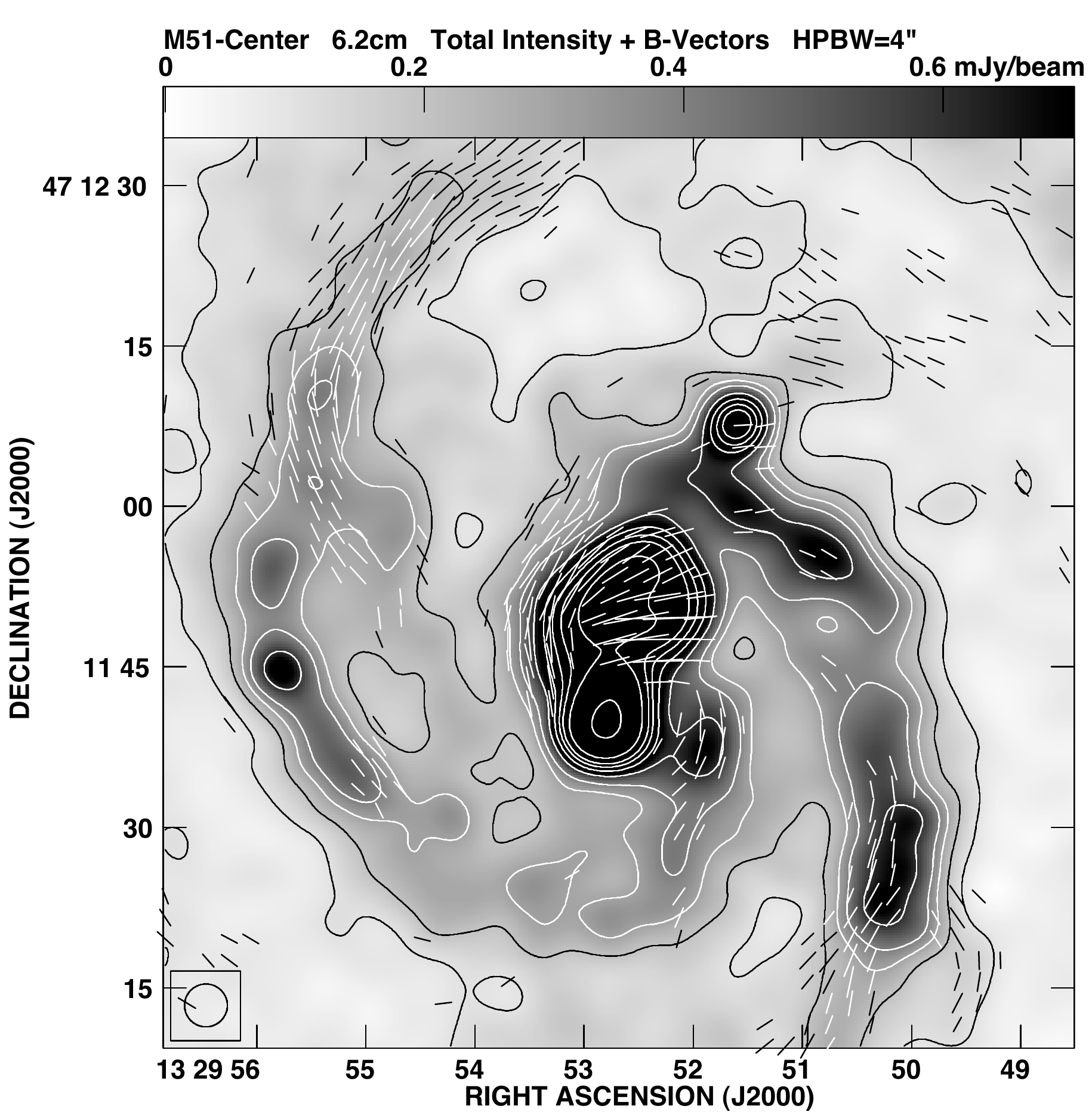}
\caption{\label{fig:inner:4b}
\wav{6} total radio emission (VLA and
Effelsberg combined) in the central $1.4\arcmin \times 1.4 \arcmin$
of M51 at $4\arcsec$ resolution.
The $B$-vectors (not corrected for Faraday rotation)
are shown where $PI\ge 3\sigma_{PI}$.}
\end{center}
\end{figure}

The polarized emission at $4\arcsec$ resolution (see
Fig~\ref{fig:inner:4b}) is also strongest along the inner arms
1--2~kpc distance from the center, with typically 20\% polarization.
The arm--interarm contrast is at least 4 in polarized intensity (this
is a lower limit as the interarm polarized emission is below the noise
level at this resolution and we take $\sigma_{PI}$ as an upper limit
for the interarm value), larger than that of the outer arms, and is
consistent with the expectations from compression of the magnetic
field in the density-wave shock (Sect.~\ref{sec:contrast}). The
contrast weakens significantly for $r<0.8 \kpc$. This may be an
indication that the inner Lindblad resonance of the inner spiral
density wave is at $r\simeq 0.8\kpc$ rather than $r\simeq 1.3\kpc$ (as
located by \citet{Elmegreen89}): the shock is probably weak around the
inner Lindblad resonance. In total intensity, the typical
arm--interarm contrast for the region of the inner arms is about 5.
The actual contrast in the M51 disc alone may be stronger than this if
there is significant diffuse emission in the central region from a
radio halo, but this effect is hard to estimate.

In the central region, two new features appear in polarized intensity which are
the brightest in the entire galaxy (Fig.~\ref{fig:inner:4b}). The first is a
region $11\arcsec$ north of the nucleus with a mean fractional
polarization of 10\% and an almost constant polarization angle. This feature
coincides with the ring-like radio cloud observed in total intensity at \wav{6}
and at $1\arcsec$ resolution by \citet{Ford85} who also detected polarization in
this region. The polarized emission indicates that the plasma cloud expands
against an external medium and compresses the gas and magnetic field. The second
feature of similar intensity in polarization is a ridge located along the
eastern edge of the first region, extending east of the nuclear source, with
15\% mean polarization and a magnetic field almost perfectly aligned along the
ridge. Field compression is apparently also strong in this ridge. The nuclear
source itself appears unpolarized at this resolution.

\section{Spectral index and magnetic field strength}
\label{sec:spx}

\begin{figure*}
\centering
\includegraphics[width=0.48\textwidth]{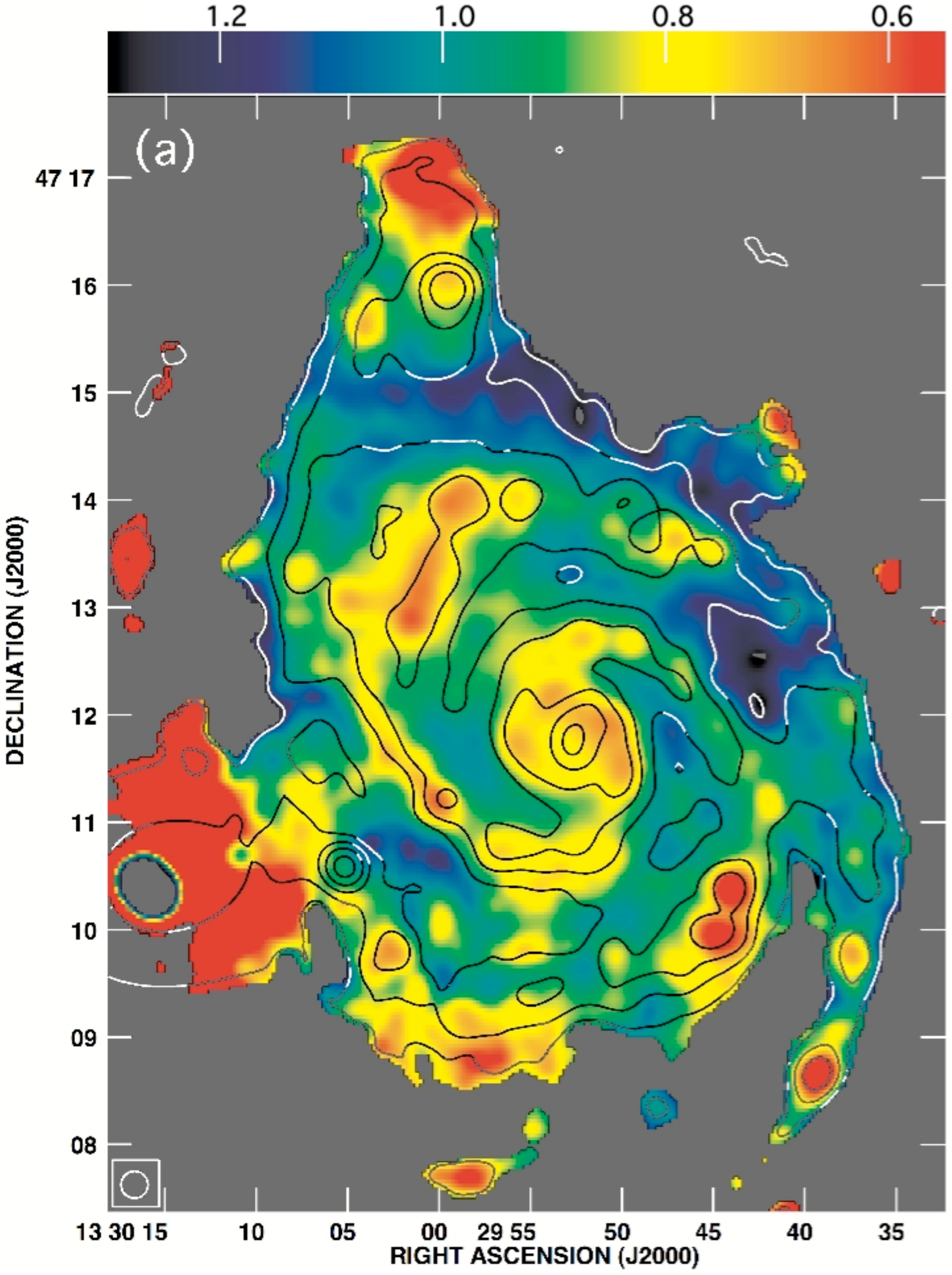}
\hfill
\includegraphics[width=0.48\textwidth]{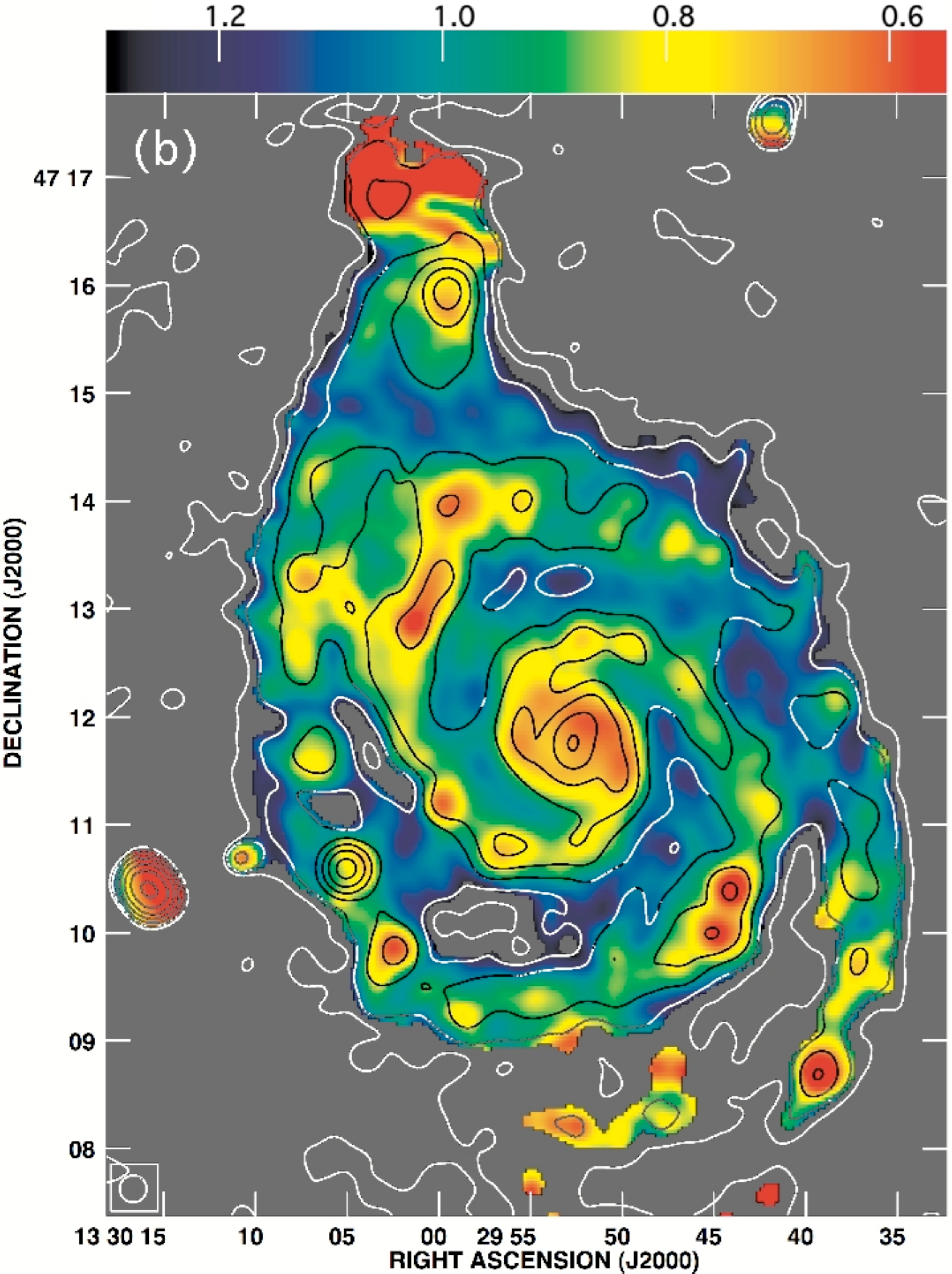}
\caption{\label{fig:spx}
The spectral index maps between \textbf{(a)} \wav{20} and \wav{3}
and \textbf{(b)} \wav{20} and \wav{6} at $15\arcsec$ resolution.
Also shown are
contours of total radio emission at \wav{3} (left) and \wav{6}
(right); the contour lines are drawn at $6,12,24,36,48,96,192$ times
the noise levels of $20\uJyb$ and $30\uJyb$ at \wav{3} and \wav{6},
respectively. Spectral indices are only calculated
where the signal at both wavelengths is $6$ times the noise level.
The error in the fitted spectral index is typically $\pm 0.02$ in
the inner galaxy and spiral arms and $\pm 0.06$ in the interarm
regions at radii $\gtrsim 1\arcmin$. The large red area at the left
edge of \textbf{(a)} is due to increased noise in the overlap region
between the two VLA-pointings at \wav{3}.}
\end{figure*}

Spectral index maps of the total radio emission at $15\arcsec$
resolution are shown in Fig.~\ref{fig:spx}. The spectral indices are
calculated using combined Effelsberg and VLA maps that contain
signal on all scales down to the
beamsize. There is therefore no missing flux in these maps due to
the short-spacings problem of interferometers.

Although there is good general agreement between the two spectral
index maps, the spectral index between \wwav{20}{3} is generally
slightly flatter than
that between
\wwav{20}{6}; this is particularly noticeable in the spiral arms.
This is due to the lower resolution of the Effelsberg \wav{6} map
(180$\arcsec$ beam
against $90\arcsec$ at \wav{3}): $180\arcsec$ is about the radius of
the M51 disc. The coarse resolution made it more difficult to
fine-tune the merging of the VLA and Effelsberg data in order to
match the integrated fluxes in the single-dish and merged maps. This
in turn led to a slight under-representation of the single-dish data
at \wav{6} in the merged map and thus to a slightly steeper spectral
index. We believe the \wwav{20}{3} spectral index, shown in
Fig.~\ref{fig:spx}a, to be more reliable.

In both spectral index maps, one can clearly distinguish arms and
interarm regions. The spectral index is typically in the range
$-0.9\le\alpha\le-0.6$ in the spiral arms and $-1.2\le\alpha\le-0.9$
in the interarm zones. Since the spectral index in the arms is not
characteristic of thermal emission ($\alpha\ne -0.1$) the arm
emission must comprise a mixture of thermal and synchrotron
radiation. The flatter arm spectral index can then be explained
by two factors: stronger thermal emission in the arms due to recent
star formation and the production of H\,{\sc ii} regions and
energy losses of cosmic ray electrons as they spread into the interarm 
from their acceleration sites in the arms.

We can use the observed steepening of the spectral index of the total
radio emission $\alpha$ to estimate the diffusion coefficient of
cosmic ray electrons in M51. We assume that the sources of the
electrons are supernovae in the arms, that the initial spectral index
is $\alpha_\mathrm{syn}\simeq -0.5$ and that the radio emission in the
interarm region is predominantly synchrotron, so that
$\alpha_\mathrm{syn}\simeq -1.1$ between the arms. A difference in the
spectral indices of $\Delta\alpha= 0.5$ is expected if the main
mechanisms of energy losses for the electrons are synchrotron emission
and inverse Compton scattering \citep{Longair:1994}.

In the next section we estimate magnetic field strengths of
around $20\uG$ for the inter-arm regions. In these magnetic fields, an
electron emitting at $5\GHz$ has an energy of about $4\,\mathrm{GeV}$.
We can estimate the lifetime of cosmic ray electrons emitting the
synchrotron radiation as
\[
\tau\simeq\frac{8.4\times 10^9}{\nu_\mathrm{16}^{1/2} \Btpe^{3/2}}\yr \approx
8.2\times 10^6 \yr, 
\] 
\citep[Sect. 1.25]{Lang:1999} where the frequency $\nu_\mathrm{16}$ is
measured in units of $16\MHz$, the total magnetic field strength in
the plane of the sky $\Btpe$ is measured in $\uG$ and we have taken
$\Btpe\simeq 15\uG$ in the interarm regions. Taking $L=1\kpc$ as the
typical distance a cosmic ray electron travels from its source in a
supernova remnant to the inter-arm region, yields the diffusion
coefficient $D$ of the electrons,
\[ D=\frac{L^2}{\tau}\simeq 4\times
10^{28}\,\cm^2\s^{-1},
\] 
which is compatible with the value of
$D\simeq1$--$10\times10^{28}\cm^2\s^{-1}$ estimated by
\citet{Strong:1998} for the Milky Way. Note that cosmic ray electrons
producing radio emission at cm wavelengths, propagating for a few
$\kpc$ in $\uG$ strength magnetic fields, give diffusion coefficients
in this range: our estimate for $D$ is not a unique property of the
cosmic rays in M51.

\subsection{Total magnetic field}

\begin{figure}
\centering
\includegraphics[width=0.45\textwidth]{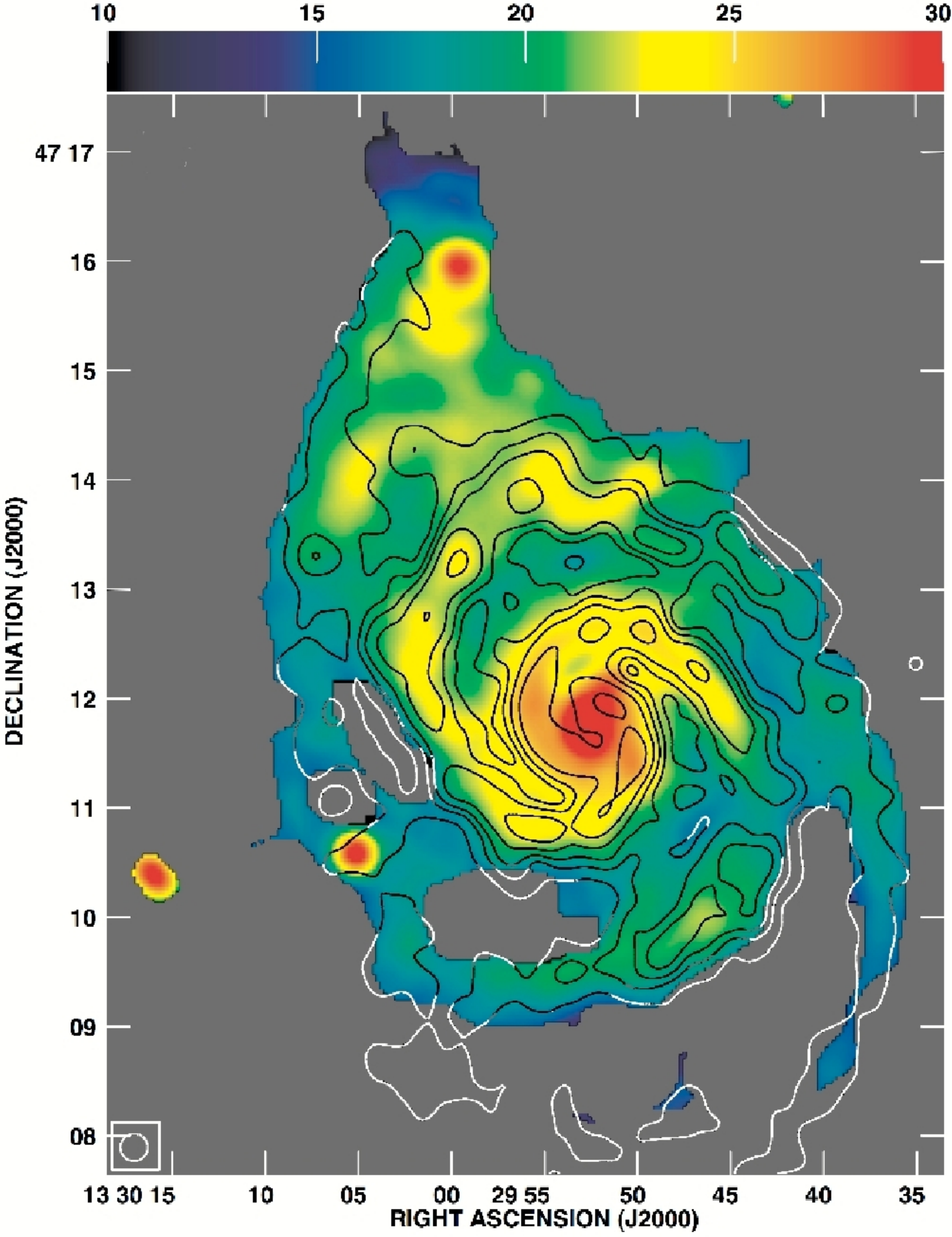}
\caption{\label{fig:Bfield}
Total magnetic
field strength, derived from the \wav{6} emission
assuming equipartition between the energy densities of magnetic
fields and cosmic rays (colour scale in $\uG$) along with
contours of neutral gas density (a combination of CO
\citep{Helfer03} and HI \citep{Rots:90} observations) plotted at
$1,4,8,16,32,64\%$ of the maximum value.}
\end{figure}

In order to derive the strength and distribution of the total magnetic
field, we have made a crude separation of the \wav{6} map at
$15\arcsec$ into its nonthermal $I_\mathrm{syn}$ and thermal
$I_\mathrm{th}$ components. We assumed that the thermal spectral index
is everywhere $\alpha_\mathrm{th}=-0.1$ as expected at cm wavelengths
\citep[e.g.][]{Rohlfs:1999} and that the synchrotron spectral index is
$\alpha_\mathrm{syn}=-1.1$ everywhere, as observed in the interarm
regions (Fig.~\ref{fig:spx}a). We are constrained in our choice of
$\alpha_\mathrm{syn}$ by two considerations: if
$\alpha_\mathrm{syn}>-1.1$ then thermal emission is absent from the
whole inter-arm region, whereas H-$\alpha$ emission is detected; if
$\alpha_\mathrm{syn}<-1.1$ we find that the degree of polarization
approaches is maximum theoretical value of $70\%$ in many regions,
which is implausible for our resolution of $570\pc$. The average
thermal emission fraction at \wav{6} is 25\%.

Assuming equipartition between the energy densities of the magnetic
field and cosmic rays, a proton-to-electron ratio of 100 and a
pathlength through the synchrotron-emitting regions of 1~kpc,
estimates for the total field strength are shown in
Fig.~\ref{fig:Bfield}, applying the revised formulae by
\citet{BeckKrause05}.

The strongest total magnetic fields of about $30\uG$ are observed in
the central region of M51. The main spiral arms host total fields of
20--$25\uG$, while the interarm regions still reveal total fields of
15--$20\uG$. This is significantly larger than in spiral galaxies like
NGC\,6946 \citep{Beck07} and M33 \citep{Tabatabaei08}. These two
galaxies have similar star-formation rates per unit area as M51, but
weaker density waves, so that compression is probably higher in M51.

\subsection{Ordered magnetic field}
\label{subsec:Breg}

The strength of the ordered magnetic field can be estimated from that
of the total field using the degree of polarization. This method gives
field strengths of $11$--$13\uG$ in the inner spiral arms,
$8$--$10\uG$ in the outer spiral arms and $10$--$12\uG$ in the
inter-arm regions. However, these values can only be attributed to a
regular (or mean) magnetic field \emph{if} the unresolved random
component is purely isotropic \citep[see][Sect.~ 5.1]{Sokoloff98}. The
observed maximum degree of polarization of around $40\%$ can equally
be produced by an anisotropic random field whose degree of anisotropy
is about 2: that is if the standard deviation of the fluctuations in
one direction on the plane of the sky is twice as large as in the
orthogonal direction. In Sect.~\ref{sec:B} we shall see that there is
only a weak signature of a regular field in the observed
multi-frequency polarization angles and that most of the polarized
emission does indeed arise due to anisotropy in the random field.

The difference between the total and ordered magnetic field strengths
gives an estimate for the isotropic random magnetic field strength of
$18\uG$, or 1.5 times the ordered field, in the arms and $13\uG$, or
1.2 times the ordered field, in the inter-arms. If the main drivers of
(isotropic) turbulence are supernova remnants, then the preferential
clustering of Type II supernovae in the spiral arms is compatible with
the higher fraction of isotropic random field in the arms. So a
significant fraction of the magnetic field consists of a random
component that is isotropic on scales less than $500\pc$.

\subsection{Uncertainties}  

Our assumption of a synchrotron spectral index that is constant across
the galaxy is crude and an oversimplification, even though the value
used,  $\alpha_\mathrm{syn}=-1.1$, can be somewhat constrained by
other data as described above. We would expect that
$\alpha_\mathrm{syn}$ should be closer to $-0.5$, the theoretical
injection spectrum for electrons accelerated in supernova remnants, in
parts of the spiral arms. This limitation results in an overestimate
(underestimate) of the thermal (nonthermal) emission in the arms.

In principle we could combine the data at all three frequencies and
simultaneously recover $I_\mathrm{syn}$, $I_\mathrm{th}$ and
$\alpha_\mathrm{syn}$ at each pixel. However we defer a more
robust calculation, interpretation and discussion to a later paper.

The equipartition estimate depends on the input parameters with a
power of only $1/(3+\alpha_\mathrm{syn}) \approx0.24$, so that even
large input errors hardly affect the results. Further errors are
induced by the underestimate of the synchrotron emission in the spiral
arms by the standard separation method (see above). In M33,
\citet{Tabatabaei:2007} found that the standard method undestimates
the average nonthermal fraction by about 25\%. In star-forming regions
of the spiral arms, the nonthermal intensity can be a factor of two
too small, which leads to an equipartition field strength which is
20\% too low. In the same regions, the synchrotron spectral index is
too steep by about 0.5 which overestimates the field strength by 15\%.
Interestingly, both effects almost cancel in M33.

The equipartition assumption itself is subject of debate.
Equipartition between cosmic rays and magnetic fields likely does
not hold on small spatial scales (e.g. smaller than the diffusion
length of cosmic rays) and on small time scales (e.g. smaller than the diffusion
time of cosmic rays).

\section{Faraday rotation and depolarization}

\subsection{Faraday rotation} 
\label{sec:rm}

\begin{figure}
\centering
\includegraphics[width=0.45\textwidth]{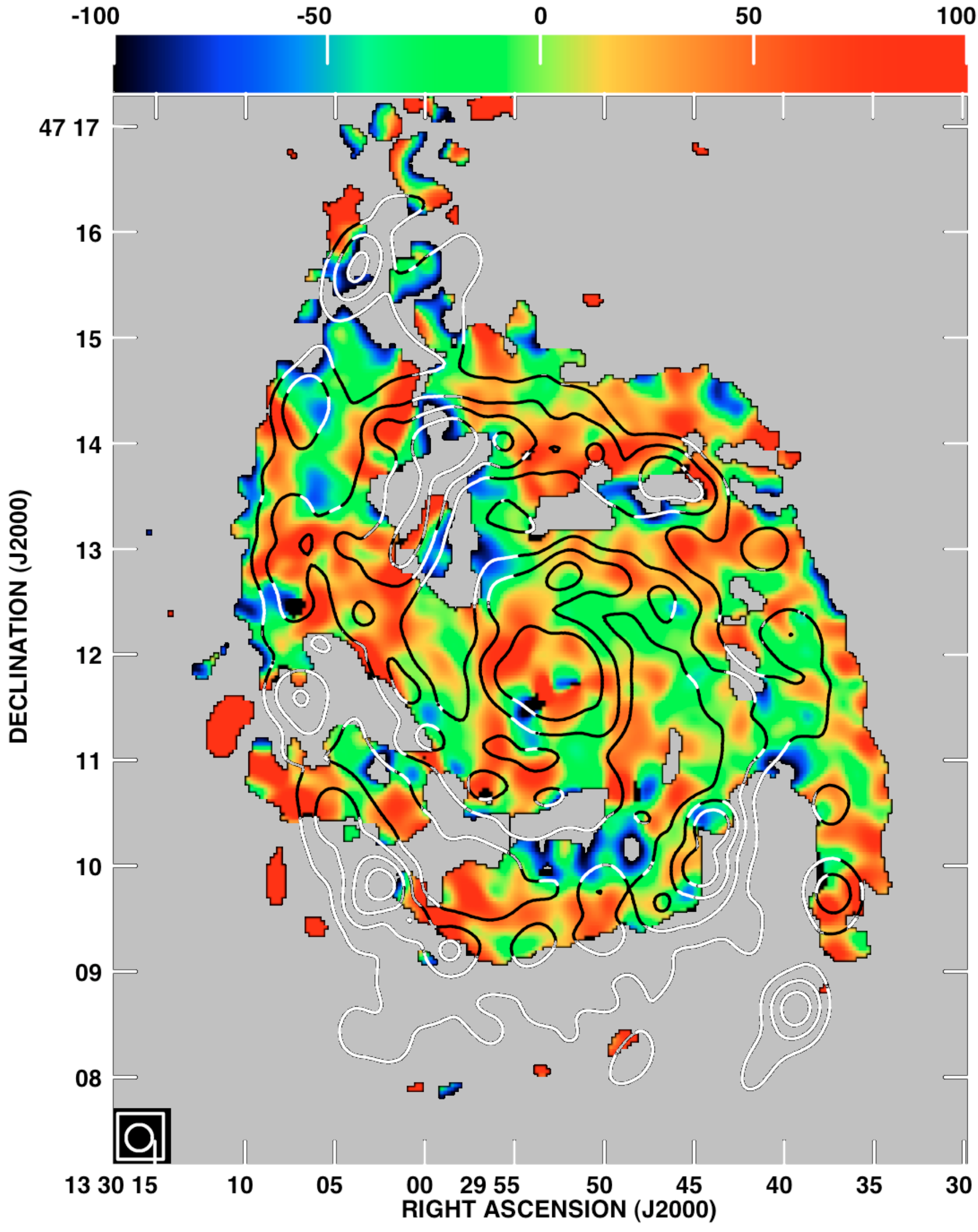}
\caption{\label{fig:rm:15}
Rotation measures between \wwav{3}{6}, at $15\arcsec$ resolution,
overlaid with contours of H$\alpha$ emission \citep{Greenawalt98} at
the same resolution, plotted at $4, 8, 16, 32\%$ of the map maximum.
Data were only used where the signal-to-noise ratio in polarized
intensity exceeds three. }
\end{figure}

\begin{figure}
\centering
\includegraphics[width=0.45\textwidth]{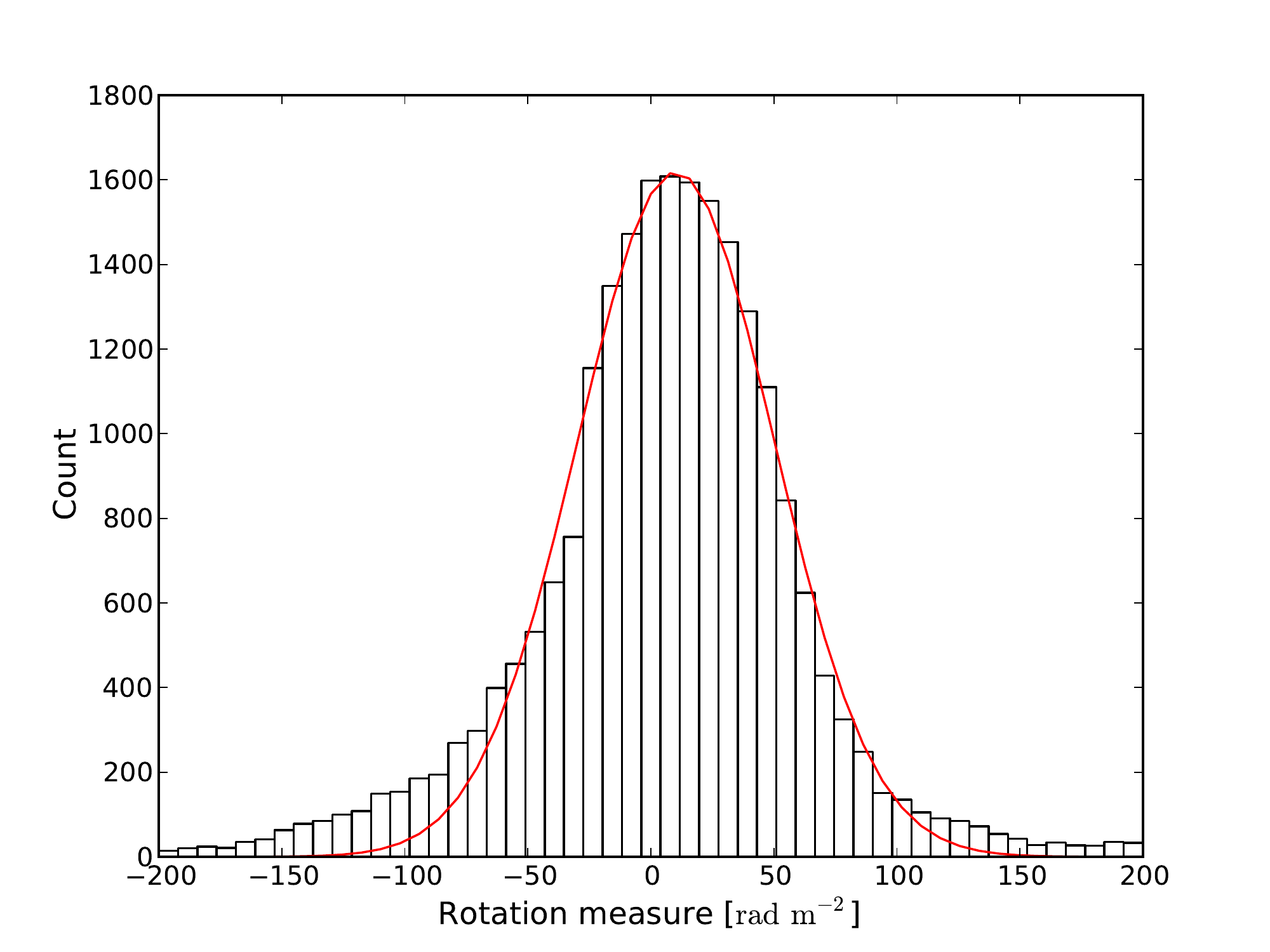}
\caption{\label{fig:rm:hist}
Distribution of rotation measures between \wwav{3}{6}, shown in
Fig.~\ref{fig:rm:15}: note that the map is oversampled and so the
histogram of pixel counts does not represent statistically independent
data points. Data were only used where the signal-to-noise ratio in
polarized intensity exceeds three. Solid line is the best-fitting
Gaussian to the histogram.}
\end{figure}

The nonthermal radio emission from the arms has a relatively low degree of
polarization (typically 25\% at \wav{6} and $15\arcsec$
resolution) so that unresolved, tangled or turbulent, magnetic field dominates
in the arms. In contrast, the interarm regions are up to $40\%$ polarized and
host a significant fraction of magnetic fields with orientation ordered at
large-scales. Whether these fields are coherent (regular) or incoherent
(anisotropic turbulent) can be decided only with the help of Faraday rotation
measures.

One might expect from the well-ordered, large-scale spiral patterns of
the polarization vectors of Figs.~\ref{fig:pi15} and \ref{fig:inner:8}
that the regular magnetic field would produce an obvious pattern in
Faraday rotation. (See the rotation measure map of M31 in Fig.~11 of
\citet{Berkhuijsen:2003} for an example of clear rotation measure
signal arising from a well-ordered magnetic field.) However, the
Faraday rotation measure map shown in Fig.~\ref{fig:rm:15} is
dominated by strong fluctuations in rotation measure, with a magnitude
of order $100\FRM$. We apparently have a paradox: the orientation of
the regular magnetic field follows a systematic spiral pattern on
scales exceeding $1\kpc$ but it does not produce any obvious large
scale pattern in $\RM$. Even the magnetic arms located between the CO
arms do not immediately exhibit any large-scale non-zero $\RM$. If the
ordered field seen in polarized intensity with an equipartition
strength of about $10\uG$ (Sect.~\ref{sec:spx}) was fully regular, we
would have $|\RM|\simeq 700\FRM$ near the major axis of M51 with a
systematic decrease moving away from this axis in azimuth, which is
not observed (we adopt $\langle n_\mathrm{e}\rangle=0.1\cmcube$,
$h=400$~pc and an inclination of $20\deg$ for this estimate, see
Sect.~\ref{sec:B:results}). Note also that the overlaid H$\alpha$
contours in Fig.~\ref{fig:rm:15} are not generally coincident with
regions of strong rotation measure.

Observational uncertainties in the measured Stokes parameters
can be a source of the fluctuations in Fig.~\ref{fig:rm:15}. 
The uncertainty in $\RM$ between \wav{3} and
\wav{6}, denoted here $\delta\RM$, depends on the signal-to-noise
ratio in the polarized intensity,
$\Sigma$, and the difference in the squares of the wavelengths
$\delta\lambda^2$ as
\begin{equation}
\delta\RM=\frac{1}{\sqrt{2}\,\Sigma\,\delta\lambda^2}\approx
\frac{280}{\Sigma} \FRM.
\end{equation}
Due to the steep spectral index of
the synchrotron emission and the
weak Faraday depolarization between \wav{3} and \wav{6}, $\Sigma$ is
lower at \wav{3} and we use these values to estimate
$\delta\RM$.
The $\RM$ fluctuations from the noise in the observed polarization signal 
are $\pm 10\FRM$ (for typical $\Sigma\gtrsim30$ in the central
$r\lesssim 90\arcsec$ of Fig.~\ref{fig:rm:15}). $\RM$ maps at $8\arcsec$
resolution have lower $\Sigma$ and are dominated by noise fluctuations,
although some strong rotation measures, above the noise level, are also present.

The structure function of the $\RM$ fluctuations is flat on scales up
to $3\arcmin$, whereas $\RM$ fluctuations due to the Milky Way
foreground in the direction of M51 in the model of \citet{Sun:2009} have a
slope of around $0.8$ (Reich, private communication). This is a strong
indication that these fluctuations are mostly due to the magnetic
field in M51.

In Fig.~\ref{fig:rm:15} the $\RM$ fluctuations with an amplitude
exceeding $45\FRM$, throughout the central $r\lesssim 90\arcsec$, and
$55\FRM$ along the outer spiral arms \emph{cannot\/} be explained by
the noise. There are therefore around ten patches where $\RM$ changes
sign over a distance of $1$--$2\kpc$ due to intrinsic fluctuations of
magnetic field.

The dispersion in $\RM$ is $15$--$20\FRM$ measured in several regions
in the inner spiral arms and $25$--$30\FRM$ in the outer arms. After
correction for the dispersion due to noise, the intrinsic dispersion
is $11\pm3\FRM$ in the inner arms and $19\pm5\FRM$ in outer arms,
hence constant within the errors. The distribution of rotation
measures is shown in Fig.~\ref{fig:rm:hist}, along with the best-fit
Gaussian, which has a mean $\RM$ of $10\pm 1\FRM$ and a dispersion of
$28\pm 1\FRM$.

Intrinsic fluctuations dominate the $\RM$ maps; even smoothed to
linear scales of order $1\kpc$, no large-scale pattern in rotation
measure is apparent, in contrast to the clear large-scale spiral
structure in polarization angles. This result is quite surprising, as
we would expect to see the components of the same field in the sky
plane and along the line of sight in polarization angle and in Faraday
rotation, respectively. As polarization angles are not sensitive to
field reversals, the observation of ordered pattern in angles does not
demonstrate the existence of a regular (coherent) field. The spiral
field seen in polarization angle could be anisotropic with many
small-scale reversals, e.g. produced by strong shearing gas motions
and compression, and hence would not contribute to Faraday rotation.
Alternatively, the field may have significant components perpendicular
to the galaxy plane (due to loops, outflows etc.) which are mostly
visible in Faraday rotation and hide the large-scale pattern. Such an
underlying large-scale pattern indeed exists, as we discuss in the
next section.

The close alignment of the observed field lines along the CO arms and
the lack of enhanced Faraday rotation in the polarized ridges can be
understood if the turbulent magnetic field is anisotropic. An
anisotropic turbulent field can produce strong polarized emission, but
not Faraday rotation. This picture is similar to that obtained for the
effect of large-scale shocks on magnetic fields in the barred galaxies
NGC~1097 and NGC~1365 \citep{Beck:2005} and will be investigated in
detail in Sect.~\ref{sec:contrast}.

\subsubsection{The size of turbulent cells}

\hl{Here we derive a new method for estimating the size of turbulent cells in the ISM of external galaxies.}

The $\RM$ dispersion $\sigma_\mathrm{RM,D}$ observed within a beam of
a linear diameter $D$ is related to $\sigma_\RM$ (Eq.~\ref{eq:sigRM})
as 
\begin{equation}\label{sRMD} \sigma_{\RM,D}\simeq
N^{-1/2}\sigma_\RM=\sigma_{\RM}\frac{d}{D}\,, 
\end{equation} 
where $N=(D/d)^2$ is the number of turbulent cells within the beam
area, assumed to be large. We confirmed the approximate scaling of
$\sigma_{\RM,D}$ with $D^{-1}$ using $\RM$ maps smoothed from
8\arcsec\ to 12\arcsec\ where the noise fluctuations are not dominant.
Combination with Eq.~(\ref{eq:sigRM}) allows us to estimate the least
known quantity involved, the diameter of a turbulent cell (or twice
the correlation scale of the turbulence):
\begin{eqnarray}
d\!\!\!&\simeq&\!\!\!
 \left[\frac{D\sigma_{\RM,D}}{0.81\langle n_\mathrm{e}\rangle
B_\mathrm{r}  (L)^{1/2}}\right]^{2/3}\\ \nonumber
&=&\!\!\! 50\pc
\left(\frac{D}{600\pc}\right)^{2/3}
\left(\frac{\sigma_{\RM,D}}{15\FRM}\right)^{2/3}
\left(\frac{\langle n_\mathrm{e}\rangle}{0.1\cm^{-3}}\right)^{-2/3}\\ \nonumber
&&\mbox{}\times\left(\frac{B_\mathrm{r}}{20\uG}\right)^{-2/3}
\left(\frac{L}{1\kpc}\right)^{-1/3}.
 \end{eqnarray}

\subsection{Faraday depolarization}
\label{sec:dp}

\begin{figure*}
\centering
\includegraphics[width=0.48\textwidth]{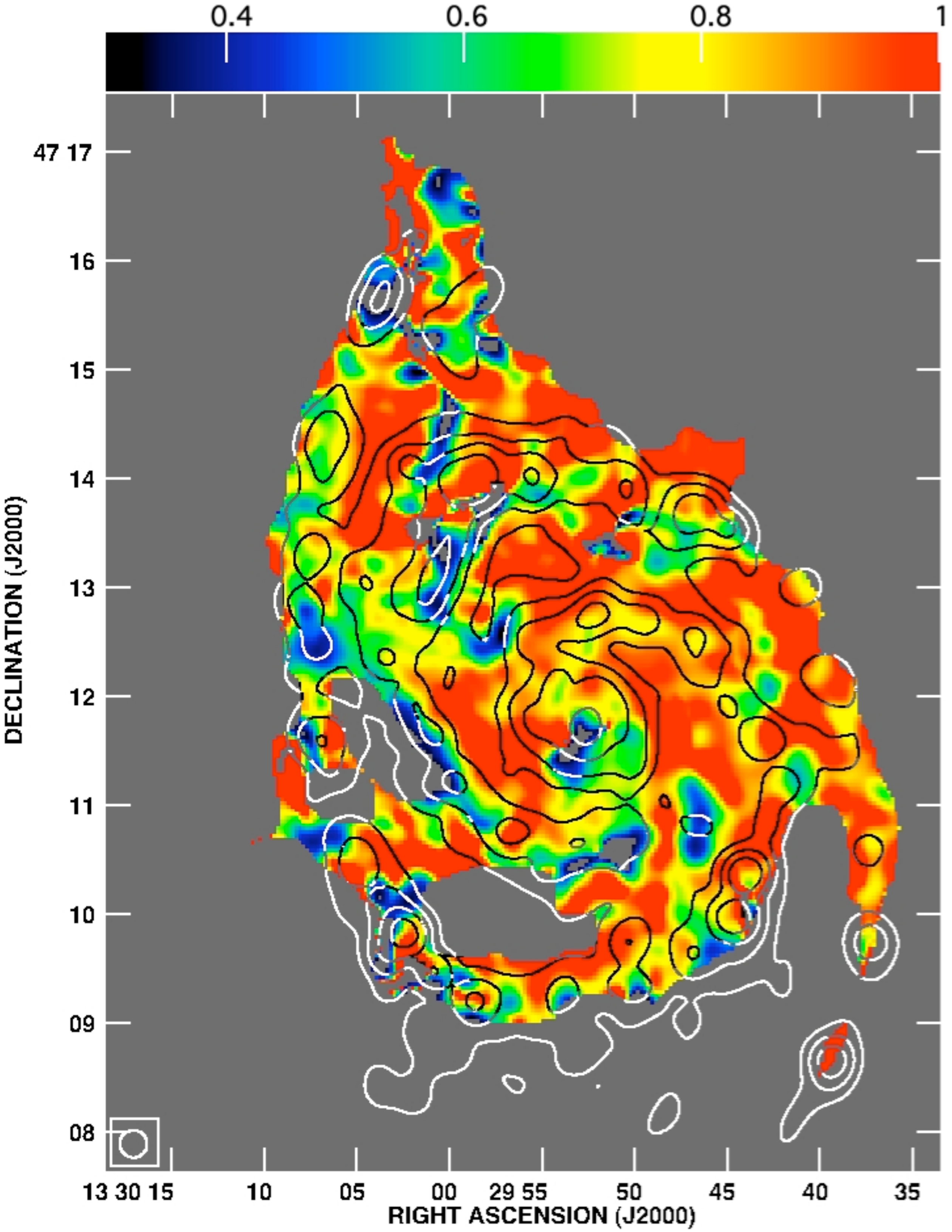}
\includegraphics[width=0.465\textwidth]{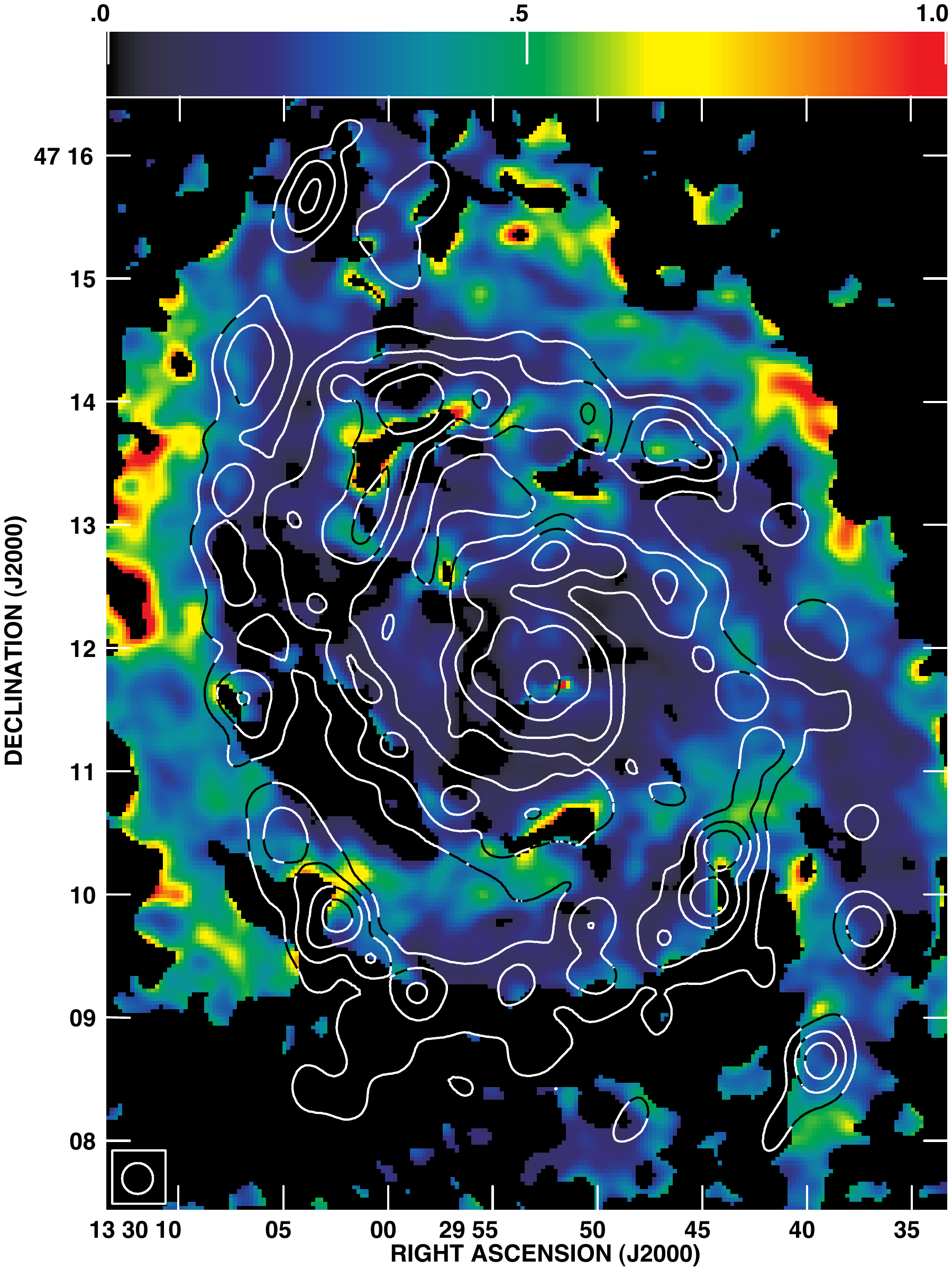}
\caption{Depolarization between \textbf{(a)} \wav{6}
and \wav{3} and \textbf{(b)} between \wav{20} and \wav{6}, both at
$15\arcsec$ resolution. Also shown are
contours of H$\alpha$ emission \citep{Greenawalt98} at the same
resolution, plotted at $4, 8, 16, 32, 74\%$ of the map maximum.}
\label{fig:dp}
\end{figure*}

Faraday depolarization $\DP$ gives important information about the
density of ionized gas, the strength of the regular and turbulent
field components, and the typical length scale (or integral scale) of
turbulent magnetic fields. $\DP$ is usually defined as the ratio of
the degrees of polarization of the synchrotron emission at two
wavelengths. This requires subtraction of the thermal emission which
is subject to major uncertainties (see Sect.~\ref{sec:spx}). Instead
$\DP$ was computed, from the polarized intensities $P$, as
$\DP=(PI_1/PI_2) \times (\nu_2/\nu_1)^{\alpha_\mathrm{syn}}$, where
$\alpha_\mathrm{syn}=-1.1$ is the synchrotron spectral index, assumed
to be constant across the galaxy. Variations in $\alpha_\mathrm{syn}$
affect $\DP$ less severely than errors in the estimate of the thermal
fraction of the total radio emission. The $\DP$ maps derived for
\wav{6} and \wav{3} and between \wav{20} and \wav{6} are shown in
Fig.~\ref{fig:dp}.

$\DP(6\cm/3\cm)$ (Fig.~\ref{fig:dp}a) is around unity (i.e.\ no
Faraday depolarization) in most of the galaxy. Small patches with 
noticeable Faraday depolarization, where $\DP=0.6$--$0.7$ are generally 
found in the spiral arms. There is no systematic connection between the
depolarization and the intensity of H$\alpha$ emission, indicating
that variations in the thermal electron density are not the main
source of the depolarization. The average value of $\DP(20\cm/6\cm)$
(Fig.~\ref{fig:dp}b) is 0.28, smaller by a factor of about 3 than
$\DP(6\cm/3\cm)$. In the inner arms, $\DP(20\cm/6\cm)$ is lower than
0.2. Only in the outer regions of the disc does $\DP(20\cm/6\cm)$
increase to 0.5 or higher.

Differential Faraday rotation within the emitting layer leads to
depolarization which varies as a $\sin(x)/x$, with $x=2 \,|\RM_i| \,
\lambda^2$, where $\RM_i$ is the intrinsic Faraday rotation measure
within the emitting layer \citep{Burn66,Sokoloff98}. With typical
values $|\RM_i| = 50\FRM$ (Fig.~\ref{fig:rm:15}), we expect little
depolarization ($\DP \approx 0.98$) at \wav{6.2} and even less at
\wav{3.6}. At \wav{20.5}, significant $\DP$ is expected for $|\RM_i| >
30\FRM$. Furthermore, lines of zero polarization (`canals') are
expected along level lines with $|\RM_i|= 37.5n\FRM$ (with integer
$n\neq0$) \citep{Shukurov:2003,Fletcher:2008}, but not a single
`canal' is observed in the \wav{20.5} polarized intensity map. This
suggests that the average $|\RM_i|$ at \wav{20.5} is significantly
smaller than $37.5\FRM$. Since the depolarization \textit{is\/}
relatively strong, it must be due to a different mechanism, e.g.,
Faraday dispersion.

Internal Faraday dispersion by turbulence in the magneto-ionic
interstellar medium
is the probable source of strong depolarization at longer
wavelengths, producing the degree of polarization given by
\citep{Burn66,Sokoloff98}:
\begin{equation}
p = p_0 \frac{1-\exp(-2S)}{2S}\, ,
\end{equation}
where $S=\sigma_\RM^2 \lambda^4$ and the maximum degree of polarization 
is $p_0\approx 0.7$. Here $\sigma_\RM$ is the
dispersion
of intrinsic rotation measure $\RM_i$ within the volume traced by
the telescope beam,
\begin{equation}
\sigma_\RM=0.81 \langle n_\mathrm{e}\rangle
B_\mathrm{r} (Ld)^{1/2}\,,
\label{eq:sigRM}
\end{equation}
where $\langle n_\mathrm{e}\rangle$ is the average thermal electron
density along the line of sight (in $\cmcube$), $B_\mathrm{r}$ the
strength of the component of the random field along the line of sight
(in $\uG$), $L$ the total path-length through the ionized gas (in pc),
$d$ the size (diameter) of a turbulent cell (in pc). Reasonable values
\hl{for the thermal disc} of $\langle n_\mathrm{e}\rangle=0.1\cmcube$, $L=800$~pc
\citep[estimated for $2.4<r<4.8\kpc$ by ][]{Berkhuijsen:1997},
$B_\mathrm{r}=20\uG$ (Sect.~\ref{sec:spx}), $d=50$~pc (see
Sect.~\ref{sec:rm}) yield $\sigma_\RM\approx 300\FRM$ resulting in
$p/p_0\approx 0.002$, a very strong depolarizing effect. The resulting
Faraday depolarization is $\DP=0.002$ between \wav{20.5} and
\wav{6.2}, much smaller than the observed depolarization of $\DP=0.2$.

\hl{We conclude that the} \wav{20} \hl{polarized emission must originate from a layer at  a greater height in the galaxy
than the bulk of the} \wav{3} \hl{and} \wav{6} \hl{polarized emission, as the}
\wav{20} \hl{disc contribution to the polarized signal that we observe in} Fig.~\ref{fig:20} \hl{must be
severely depolarized.} Two possibilities, that are not mutually
exclusive, are that (i) the scale height of the \wav{20} synchrotron
disc is greater than the scale height of the thermal electron disc, or
(ii) the \wav{20} polarized emission is produced in a synchrotron
halo. In either case, since Faraday rotation is observed at
wavelengths around \wav{20} \citep{Horellou:1992, Heald:2009}, with
about $1/5$ the amplitude as between \wav{3} and \wav{6} (see below),
thermal electrons and a magnetic field are required in the
halo.\footnote{The terminology used here is unavoidably imprecise.
Instead of a disc and halo we could just as easily refer to a thin and
thick disc. We are unable to say anything about the geometry (e.g.
flat or spheroidal) of the two layers using our data, only that there
must be two layers producing the Faraday rotation observed in M51. In
Section~\ref{sec:B} we look at the magnetic field structure in the two
layers in more detail.}

In contrast, at \wav{3.6} $\sigma_\RM\approx 300\FRM$ produces
virtually no depolarization with $p/p_0\approx 0.85$, so we expect
that the galaxy is transparent (or Faraday thin) at \wav{3.6}. At
\wav{6.2} we have $p/p_0\approx 0.3$, moderate depolarization, with
$DP=0.2$ between \wav{6.2} and \wav{3.6}. This rough estimate for $DP$
is the same order of magnitude as the $DP$ shown in
Fig.~\ref{fig:dp}(a), albeit about three times lower than the typical
observed value of $DP\approx 0.9$, indicating that our value of
$\sigma_\RM$ is a slight overestimate. (The uncertainty in the adopted
values of $\langle n_\mathrm{e}\rangle$, $L$ and $B_\mathrm{r}$ can easily 
explain the discrepancy: for $\sigma_\RM\approx 200\FRM$ we have 
$DP=0.6$, much closer to the observed values.) Since
Faraday effects are so small at \wav{3.6}, Fig.~\ref{fig:dp}(a)
strongly indicates that the disc is also Faraday thin at \wav{6.2}.

\section{Regular magnetic field structure}
\label{sec:B}

\subsection{The method}

The map of Faraday rotation discussed in Sect.~\ref{sec:rm} clearly
shows the effects of strong magnetic field fluctuations on scales of
$300\pc$ to $\sim 1\kpc$; from this map one could expect the
magnetic field to be rather chaotic and disordered. However, the
observed polarization angles, shown in Figs.~\ref{fig:i15} and
\ref{fig:pi15}, suggest an underlying spiral pattern to the magnetic
field on scales $\gtrsim 1\kpc$ even when corrected for Faraday
rotation (Fig.~\ref{fig:inner:co}). If the large-scale spiral pattern
in the magnetic field is due to a regular field component, such
as might be expected due to mean-field dynamo action, we can expect
to find a signature of such a field in the Faraday rotation
signal at relevant scales.  If no such signal can be uncovered then
we may conclude that the data do not support the presence of a mean
field in M51 and that the spiral patterns seen in Figs.~\ref{fig:i15}
and \ref{fig:pi15} are being imprinted on a purely random magnetic field
through e.g. large-scale compression in the gas flow.

\begin{figure}
\centering
\includegraphics[width=0.45\textwidth]{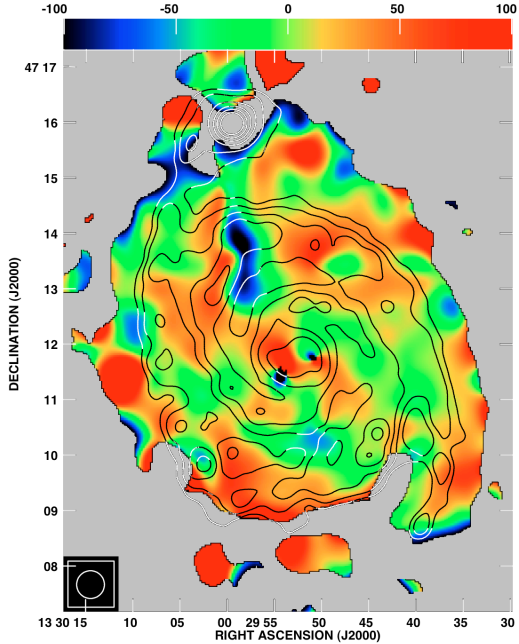}
\caption{\label{fig:rm:30}
Rotation measures between \wwav{3}{6},
at $30\arcsec$ resolution, overlaid with contours of mid-infrared
15$\,\mu$m emission \citep{Sauvage96} plotted at $1,2,4,8,16,32\%$
of the maximum value.
Data were only used where the
signal-to-noise ratio in polarized intensity exceeds five. }
\end{figure}

\hl{Figure}~\ref{fig:rm:30} \hl{shows a Faraday rotation map using our} \wav{3} \hl{and} \wav{6} \hl{data smoothed to $30\arcsec$ resolution: large-scale structure in the rotation measure distribution across the disc is now visible (cf. Fig.}~\ref{fig:rm:15}). \hl{The rotation measures between} \wav{18} \hl{and} \wav{20} \hl{also show large-scale structure} \citep[Fig. 9]{Horellou:1992}, \hl{but with a markedly different pattern. The complex regular magnetic field structure producing the observed rotation measure patterns cannot be reliably determined from an intuitive analysis of these maps. For example, a simple axisymmetric or bisymmetric field would produce a single or double period variation with azimuth, whereas the actual pattern is clearly more complicated. A particular difficulty is that the pattern in the observed Faraday rotation is very different between the pairs of short and long wavelengths, indicating that two Faraday-active layers may be present.}  

In order to look for the signature of a regular magnetic field in the
multi-frequency polarization maps we applied a Fourier filter to the
$15\arcsec$ resolution Stokes $Q$ and $U$ maps at \wwwav{3}{6}{20} to
remove the signal on scales $\lesssim 30\arcsec \approx 1.1\kpc$. We
also used the \wav{18} data of \citet{Horellou:1992} which has a
resolution of $43\arcsec$ resolution and was not filtered. Then maps
of polarization angle were constructed at each wavelength. These were
subsequently averaged in sectors with an opening angle of $20\deg$ and
radial ranges of $2.4-3.6\kpc$, $3.6-4.8\kpc$, $4.8-6.0\kpc$ and $6.0-7.2\kpc$
(see Fig.~\ref{fig:Bmodel}). We estimated that the minimum systematic
errors in polarization angle arising from this method are about
$4\degr$ at \wwav{3}{6} and $10\degr$ at \wwav{18}{20}, with the main
source of error being uncertainties arising from Faraday rotation by
the random magnetic field \citep[see][Sect.~2]{Ruzmaikin:1990}. We set
these as minimum errors in the average polarization angles, otherwise
using the standard deviation in the sector.

We applied a method that seeks to find statistically good fits to the
polarization angles using a superposition of azimuthal magnetic field
modes $\exp(im\phi)$ with integer $m$, where $\phi$ is the azimuthal
angle in the galaxy's plane measured anti-clockwise from the north end
of the major axis. A three-dimensional model of the regular magnetic
field is fitted to the observations of polarization angles at several
wavelengths simultaneously. The polarization angle affected by Faraday
rotation is given by $\psi=\psi_0+\RM\lambda^2+\RMfg\lambda^2$, where
the intrinsic angle of polarized emission is $\psi_0$, $\RM$ is the
Faraday rotation caused by the magneto-ionic medium of M51 and $\RMfg$
is foreground Faraday rotation arising in the Milky Way.  The method
of modelling is described in detail in \citet{Berkhuijsen:1997} and
\citet{Fletcher:2004} and has been successfully applied to normal
spiral galaxies \citep{Berkhuijsen:1997, Fletcher:2004, Tabatabaei08}
and barred galaxies \citep{Moss:2001, Beck:2005}. We have derived a
new model of the regular magnetic field in M51 as our combined VLA +
Effelsberg maps at \wwav{3}{6} are a significant improvement in both
sensitivity and resolution over the maps used by
\citet{Berkhuijsen:1997}.

The fitted parameters of the regular magnetic fields are given in
Appendix~\ref{app:Bmodel}. These fit parameters can be used to
reconstruct the global magnetic structure in M51. In order to obtain
statistically good fits to the observed data we needed plane-parallel
field components only (no vertical fields). No satisfactory fits could
be found, using various combinations of the azimuthal modes m=0, 1, 2
for the horizontal field and m=0,1 for the vertical field, for the
observations at all 4 wavelengths for a single layer. Therefore at
least two separate regions of Faraday rotation are required. This is
because the patterns of polarization angle and Faraday rotation at
\wwav{3}{6} and \wwav{18}{20} are very different: at \wav{18} and
\wav{20} the disc emission is heavily depolarized by Faraday
dispersion (see Sect.~\ref{sec:dp}), so we only see polarized emission
at these wavelengths from the top of the disc. A similar requirement
for two Faraday rotating layers in M51 was also found by
\citet{Berkhuijsen:1997}.

\begin{figure}
\centering
\includegraphics[width=0.49\textwidth]{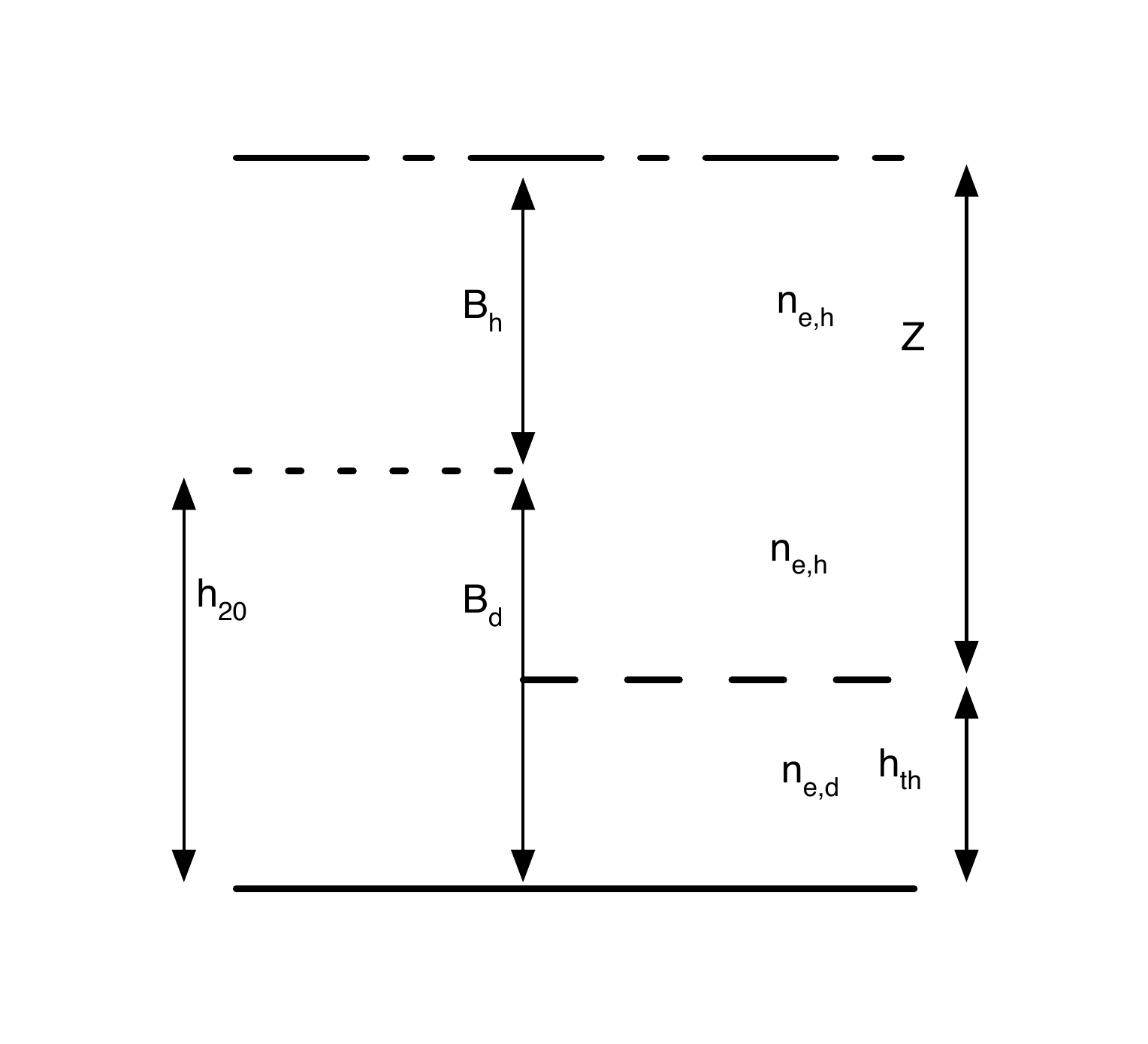}
\caption{Geometry of the disc and halo layers. The thermal disc and
halo have thickness $\hth$ and $Z$, and electron densities
$n_{\el,\mathrm{d}}$ and $n_{\el,\mathrm{h}}$. The scale-height of the
region emitting polarized synchrotron radiation at \wwav{18}{20} is
$h_{20}$ and the \emph{observed} \wav{20} polarized emission comes
from the layer $\hth<z<h_{20}$, due to Faraday depolarization. The
regular magnetic field has two layers: the disc $B_\mathrm{d}$,
extending to the same height as $h_{20}$, which is both an emitting
and Faraday rotating layer, and the halo $B_\mathrm{h}$ where only
Faraday rotation occurs.}
\label{fig:layers}
\end{figure}

To describe the two layers in the model, the Faraday rotation from M51
is split into two components, arising from a disc and halo,
$\RM=\xi^{(D)}\RM^{(D)}+\xi^{(H)}\RM^{(H)}$, where $\xi^{(D)}$ and $
\xi^{(H)}$ are parameters that allow us to model how much of the disc
and halo are visible in polarized emission at a given wavelength. We
use this decomposition of the $\RM$ into disc and halo contributions
to take into account the depolarization of the \wav{20} emission from
the thermal disc discussed in Sect.~\ref{sec:dp} by setting
$\xi^{(D)}=0$ and $\xi^{(H)}=1$ at \wwav{18}{20} and $\xi^{(D)}=1$ and
$\xi^{(H)}=1$ at \wwav{3}{6}. In other words, at \wwav{18}{20} the
polarized emission is produced in a thin layer that lies above the
thermal disc (see Sect.~\ref{sec:dp}) and has the same regular
magnetic field configuration that produces the \wwav{3}{6} polarized
emission. The \wwav{3}{6} emission is Faraday rotated in the thermal
disc and the halo whereas the \wwav{18}{20} emission is only Faraday
rotated in the halo: see Fig.~\ref{fig:layers}.

\subsection{Results}
\label{sec:B:results}

\begin{figure*}
\centering
\includegraphics[width=0.49\textwidth]{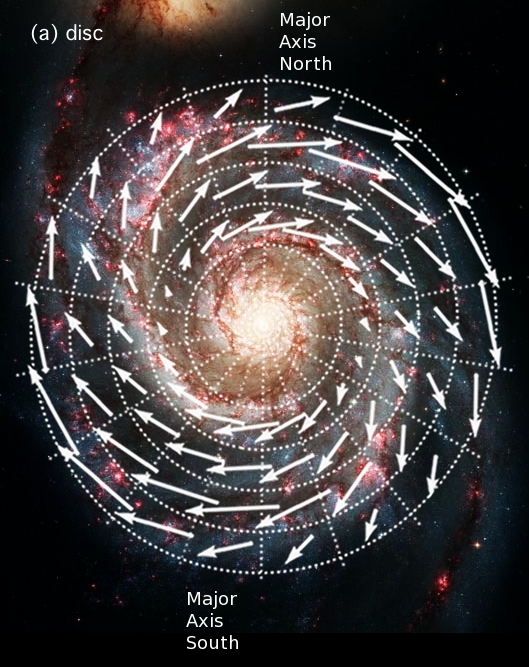}
\hfill
\includegraphics[width=0.49\textwidth]{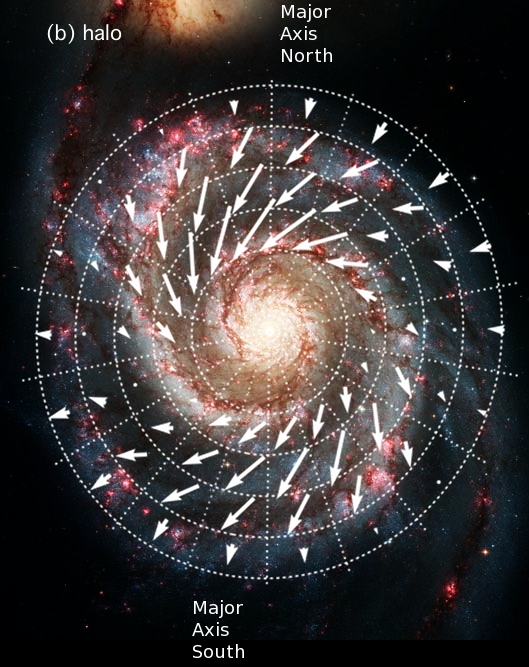}
\caption{\label{fig:Bmodel} The regular magnetic field derived from
fitting a model to the observed polarization angles at
\wwwwav{3}{6}{18}{20}, with the length of the magnetic field vectors
proportional to the field strength, overlaid on the same optical image
as in Fig.~\ref{fig:i15}. \textbf{(a)} The regular
magnetic field in the galactic disc. \textbf{(b)} The regular
magnetic field in the galactic halo. Ring boundaries are at $2.4$,
$3.6$, $4.8$, $6.0\kpc$ and $7.2\kpc$ and all sectors have an opening angle of
$20\deg$. The major axis is indicated: the midpoints of these two
sectors correspond to $\phi=0\deg$ and $\phi=180\degr$ respectively. }
\end{figure*}

The resulting regular magnetic field structure is shown in
Fig.~\ref{fig:Bmodel}. The regular field in the disc is best described
by a superposition of m=0 and m=2 horizontal azimuthal modes and has a
radial component directed outwards from the galaxy centre, whereas the
halo field has a strong $m=1$ horizontal azimuthal mode and is
directed inwards in the north, opposite to the direction of the disc
field, and outwards in the south, same as the disc field. Oppositely
directed components of the field in the disc and halo were also found
by \citet{Berkhuijsen:1997}, however our new observations place the
strong $m=1$ mode, and the resulting strong asymmetry in the magnetic
field, in the halo rather than the disc. We believe that the
difference in fitted fields arises from the higher quality of the new
\wwav{3}{6} data, as discussed above. \citet{Heald:2009} derived
rotation measure maps from multi-channel WSRT data at \wav{22} using
the RM-synthesis technique that qualitatively show the pattern
expected from a $m=1$ magnetic field. Since depolarization due to
Faraday dispersion cannot be removed by RM-synthesis, this gives a
second, independent, indication that the halo of M51 hosts an $m=1$
regular magnetic field.

In the ring $2.4<r<3.6\kpc$ a weak $m=0$ mode is required to fit the
data. This arises due to a sharp change in the observed polarization
angles at \wwav{18}{20} between $\phi=120\degr$ and $\phi=140\degr$.
Even with two halo modes we cannot capture the rapid change in angles
and had to exclude the \wav{20} data at $\phi=140\degr$: the
alternative would be to add an extra mode, with three new parameters,
to model one data point.

The process by which two different regular magnetic field patterns in
two layers of the same galaxy are produced is not clear and is beyond
the scope of this paper. We only offer some speculative suggestions:
(i) both fields might be generated by mean field dynamos but operating
in different regimes, with the interaction of M51 with NGC~5195
driving a $m=1$ mode in the halo; (ii) the halo field could be a relic
of the magnetic field present in the tenuous intergalactic medium from
which the galaxy formed; (iii) as the disc field is advected into the
halo it can be modified by the halo velocity field into the $m=1$
pattern. All of these possibilities will require careful modelling to
determine their applicability to the problem.

In the disc of M51, in the four rings used in our model the $m=0$
azimuthal field component is $1$--$2$ times the strength of the
$m=2$ mode (Table~\ref{tab:fit}). While the strength of the $m=2$ mode remains
approximately constant between the rings, the $m=0$ mode is of
similar strength in the inner ring, but is much stronger in the
other three rings.

The r.m.s. regular field strength $\bar{B}$ in each ring can be
determined by integrating the fitted modes over azimuth (Table~\ref{tab:fit}) via
\[
\bar{B} = \frac{\bar{R}}{88\FRM}\left(\frac{\bra{n_\mathrm{e,d}}}{0.11\cmcube}\right)^{-1}
\left(\frac{h_{\mathrm{th}}}{1\kpc}\right)^{-1}\, \uG,
\]
where
\[
\bar{R}=\frac{1}{2\pi}\int^{2\pi}_{0}\sqrt{R_r^2 + R_{\phi}^2}\,\mathrm{d}\phi
\]
with $R_r$ and $R_{\phi}$ given by Eq.~(\ref{eq:Bmod}).
\citet{Berkhuijsen:1997} estimated that $\bra{n_\mathrm{e,d}}=0.11\cmcube$
and $\hth=400\pc$ in the radial range\footnote{\citet{Berkhuijsen:1997} adopted
a distance to M51 of $8\Mpc$: we have rescaled their radii to our distance
of $7.6\Mpc$.} 
$2.4<r<4.8\kpc$ and for $4.8<r<7.2\kpc$ they
estimated $\bra{n_\mathrm{e,d}}=0.06\cmcube$ and $\hth=600\pc$. The
r.m.s. strengths of the large-scale magnetic field,
using these parameters, are $\bar{B}=1.4\pm0.1\uG$, $\bar{B}=1.7\pm0.5\uG$,
$\bar{B}=2.7\pm1.0\uG$ and $\bar{B}=2.8\pm0.1\uG$ in the rings $2.4<r<3.6\kpc$,
$3.6<r<4.8\kpc$, $4.8<r<6.0\kpc$ and $6.0<r<7.2\kpc$ respectively. 
These are a factor
of 4 lower than the strength of the ordered field derived from
the equipartition assumption (Sect.~\ref{subsec:Breg}). The
equipartition field strength is based on the observed polarized
intensity and an anisotropic random magnetic field can contribute to
the polarized signal (see Sect.~\ref{subsec:Breg}) but will not produce
any systematic pattern in
polarization angles at different frequencies. This is probably the
reason for the discrepancy:
most of the polarized radio emission in M51 does not trace a mean magnetic
field, only the modelled large-scale pattern in Faraday rotation does.

The average pitch angle of the $m=0$ mode is $-20\deg$ with little
variation in radius between the rings. This means that the spiral
structure of the regular field is coherent over the whole galaxy. The
weaker $m=2$ modes produce an azimuthal
variation in pitch angle of about $15\degr$ in the inner ring and
$5\degr$ in the other rings. This variation of the pitch angle of
the \emph{regular} magnetic field with azimuth is much lower than
the variation of the pitch angle of the \emph{ordered} magnetic
field observed in polarization: \citet{Patrikeev06} showed that the
orientation of Faraday rotation corrected polarization angles
change by about $30\degr$. Thus the anisotropic random magnetic
field, that we believe produces most of the polarized emission in
M51, has a stronger azimuthal variation in its orientation: this is
to be expected if the anisotropy arises from some periodic mechanism
such as compression in spiral arms or localised enhanced shear.

The bisymmetric halo field has a much larger pitch angle of
about $-50\deg$ in the inner three rings. If this field is generated
by a mean field dynamo the high pitch angles may be an indication
that differential rotation in the halo is weak and an
$\alpha^2$-dynamo action is significant. (In an $\alpha^2$-dynamo,
shear due to differential rotation is negligible and so
$|B_r|\approx|B_{\phi}|$.) The average thermal electron density and
size of the halo in this galaxy are unknown and we have no specific
constraints to apply. Taking reference values of
$\bra{n_\mathrm{e,h}}=0.01\cmcube$, $Z=5\kpc$ and
$\bra{n_\mathrm{e,h}}=0.06\cmcube$, $Z=3.3\kpc$ for the radial ranges
$2.4\le r\le4.8\kpc$ and $4.8\le r\le7.2\kpc$ respectively, where the
densities are one tenth of the disc density as in the Milky Way and
$Z$ is as estimated by \citet{Berkhuijsen:1997}, the r.m.s. strength
of the halo field (note that the $m=1$ mode means there are two
values of the azimuth in each ring where the field is zero) is
\[
\bar{B}\approx \frac{\bar{R}_{h}}{8\FRM}\left(\frac{\bra{n_\mathrm{e,h}}}{0.01\cmcube}\right)^{-1}
\left(\frac{Z}{1\kpc}\right)^{-1}\, \uG,
\]
where $\bar{R_{h}}$ is the average amplitude of the halo mode.
The fitted amplitudes of the halo field given in Table~\ref{tab:fit}
give r.m.s. regular field strengths in the halo of $1.3\pm0.3$,
$1.2\pm0.4$, $2.2\pm1.0$ and $1.6\pm0.6\uG$ in the four rings, with
increasing radius.

We are confident that the azimuthal modes and pitch angles, fitted to
our data, for the regular magnetic field in the disc and halo are
robust as we have carried out extensive checks and searches of the
parameter space. For example, if the \wwav{18}{20} data is ignored and
a disc only model used the \wwav{3}{6} data produce a very similar
fitted field to that given in Table~\ref{tab:fit}. However, we cannot
consider the mode amplitudes to be reliable other than to reflect the
relative strengths of the $m=0$ and $m=2$ modes in the disc.

A field reversal between the regular fields in the disc and the
inner halo has also been suggested in the Milky Way \citep{Sun08},
perhaps similar to the northern half of M51
(Fig.~\ref{fig:Bmodel}). 

It is probable that our model for the vertical structure of M51
(Fig.~\ref{fig:layers}) is too simple and that this has lead to too
much Faraday rotation being put into the halo field at the expense of
the disc. In particular we do not allow for Faraday rotation from the
\hl{thin} emitting part of the \wav{20} disc nor for emission at any wavelength
from the halo. Furthermore, each azimuthal mode is assumed to have an
azimuth-independent intrinsic pitch angle. Adding more parameters to
this model, given the limited number of sectors and frequencies that
we can use, will not likely resolve this question. A more productive
approach will be to develop a new model that also takes into account
depolarizing effects directly and whose outputs are statistically
compared directly to the individual maps, including the unpolarized
emission (perhaps using the maps of Stokes parameters themselves).

\section{Arm--interarm contrasts}
\label{sec:contrast}

\begin{table*}
\caption{\label{con}Arm--interarm contrasts in observed
quantities. Radio intensities at \wav{6} are shown. The data has been
smoothed to $15\arcsec$ resolution and the arm and interarm regions
identified using a mask derived from the wavelet transform of the
combined CO and HI map (see the text for details). The inner arm
data are for a resolution of $4\arcsec$ at a position of strong
contrast in polarized radio emission.}

\begin{center}
\begin{tabular}{lccccccc}
\hline
Quantity & Units & \multicolumn{3}{c}{Average $1.6\kpc<r<4.8\kpc$} & \multicolumn{3}{c}{Inner arms $0.8\kpc<r<1.6\kpc$}\\
        & & Arm & Interarm & Arm/interarm ratio & Arm & Interarm & Arm/interarm ratio \\
\hline
Neutral gas column density & $10^{21}\mathrm{H}\cm^{-2}$ & 18.0 & 4.0 & 4.5 & 200 & 40 & 5 \\
Total radio intensity & $\mJyb$ & 1.1 & 0.5 & 2.2 & 0.6 & 0.12 & 5 \\
Polarized radio intensity & $\mJyb$ & 0.1 & 0.1 & 1.0 & 0.07 & $<0.01$ & $\ge 7$\\
\hline
\end{tabular}
\end{center}
\label{tab:arminter}
\end{table*}%

We wish to investigate the effect of magnetic field compression in
the large-scale shocks associated with the spiral arms. In
particular, how the regular and random magnetic field components are
changed by the shocks and whether shock compression of isotropic
random magnetic fields can produce enough anisotropic field to
account for most of the polarized emission as inferred in
Sections~\ref{sec:dp} and \ref{sec:B}. We have carefully examined the azimuthal variations in
the various maps at different radii: these emission profiles clearly demonstrate
that there is not a simple azimuthal behaviour of any of the
measured quantities. One cannot identify ``typical'' arm to
interarm contrasts. So we have used a mask to separate arm (pixels
in the mask set to $1$) and interarm (mask pixels set to $-1$)
regions in each of the maps and hence calculate the average contrast
over a wide radial range.

We combined the CO map of \citet{Helfer03} with the HI map of
\citet{Rots:90} to produce a map of the total neutral gas density at
$8\arcsec$ resolution, assuming a constant conversion factor
$N_{\mathrm{H}_2}=1.9\times 10^{20} I_\mathrm{CO}\,\cm^{-2}$. 
The arm--interarm mask was determined by making a wavelet transform, using the
Mexican hat wavelet with a linear scale of about $1\kpc$, of this map. 
This scale was selected by
examining a range of transform maps: a $1\kpc$ scale wavelet
produces transform coefficients that are continuously positive along
spiral arms and continuously negative in interarm regions. Also,
$1\kpc$ seems reasonable as a typical width of the spiral arms in
M51. This mask was used to separate the arm and interarm components
of each of the maps listed in Table~\ref{tab:arminter}, over the
radial range $1.6\kpc<r<4.8\kpc$, and then the average arm and interarm
values were determined.

In addition to the data in Table~\ref{tab:arminter} we also separated
the rotation measures shown in Fig.~\ref{fig:rm:15} into arm and
interarm components. We calculated the average magnitude
$\left<|\RM|\right>$ and standard deviation $\sigma_\RM$ of $\RM$ and
found that $\left<|\RM|\right>\approx 22\FRM$ and $\sigma_\RM\approx
39$ in the interarm and $\left<|\RM|\right>\approx 15\FRM$ and
$\sigma_\RM\approx 29$ in the arms. This may indicate that the regular
magnetic field is stronger in the interarm regions than in the arms.
However, the interpretation is difficult as the $\RM$ distribution
depends on the signal-to-noise ratio, which tends to be higher in the
arms.

The
contrast in the neutral gas column density (2H$_2$ + HI) is compatible
with what might be expected from compression by a strong adiabatic
shock,
\[
\epsilon_n=n\down/n\up=4,
\]
where the superscripts  $\down$ and $\up$ refer to downstream and
upstream of the shock front. We note that the scale
height of the gas layer $h$ is not expected to be much
affected by the spiral pattern \citep{Shukurov98}; H\,{\sc i}
observations in the Milky Way suggest $h\down/h\up\simeq1$--$1.5$ in
the outer Milky Way \citep{K07}. 

\subsection{General considerations}
Explaining the arm--interarm contrast in the observed radio intensity
is a long-standing problem. \citet{Mathewson72} argued that their
\wav{20} observations of M51 are consistent with the density wave
theory of \citet{Roberts:1970}. They assumed that both cosmic ray
number density and the tangential magnetic field increase in
proportion to the gas density at the spiral shock. The resulting
arm--interam contrast in radio intensity, after taking into account of
the telescope beamwidth, is expected to be of order $10$ or more.

\citet{Tilanus:1988}, using observations at a higher resolution, found
that the shape of cross-sectional profiles across the radio intensity
arms is not compatible with the density wave theory and concluded that
the synchrotron emitting interstellar medium is \emph{not} compressed
by shocks and decouples from the molecular clouds as it traverses the
arms. Thus there is clearly a discrepancy between the physically
appealing theory of \citet{Roberts:1970} and observations \citep[see
also][p. 590]{Condon:1992}.

\citet{Mouschovias:1974} and \citet{Mouschovias:2009} suggest that
only a moderate increase in synchrotron emission in spiral arms is
expected due to the Parker instability: rather than being strongly
compressed the regular magnetic field rises out of the disc, in loops
with a scale of 500--1000~pc. However, the substantial random
component of the magnetic field in M51 may supress the instability or
reduce it to a simple uniform buoyancy \citep{Kim:2001}. Furthermore
we do not observe the periodic pattern of enhanced Faraday rotation
along the spiral arms that would be expected from the vertical
magnetic fields at the loop footpoints (Fig.~\ref{fig:rm:15}).

In this section we reconsider this question with additional emphasis
on the polarized intensity. In Sect.~\ref{subsec:cr}, we consider how 
shock compression affects
the emission of cosmic-ray electrons at a \emph{fixed} frequency. The
effect of the modest observational resolution on the arm--interarm
contrasts is discussed in Sect.~\ref{subsec:beam}. In
Sect.~\ref{subsec:comp} we consider the case of compression of an
isotropic random magnetic field and conclude that this may be the
dominant origin of the arm--interam contrast in radio intensity only
in the inner galaxy. Finally, in Sect.~\ref{subsec:decomp} we show
how the decompression of an isotropic random magnetic field as it
leaves the spiral arm affects the arm--interarm contrast.

\subsection{Cosmic rays in compressed gas}
\label{subsec:cr}

Assuming that magnetic field is parallel to the shock front of the spiral density wave and is
frozen into the gas, its strength increases in proportion to the gas
density, $B\propto\rho$, as appropriate for one-dimensional
compression. The ultrarelativistic gas of cosmic rays, whose  speed of
sound is $c/\sqrt3$, with $c$ the speed of light, is not compressed in
the arms. However, compression of magnetic field will affect the
cosmic rays (including the electron component) because
$p_\perp^2/B\approx\const$ is an adiabatic invariant
\citep{Rybicki:1979}, where $p_\perp=\gamma m\el c$ is the component
of the particle momentum perpendicular to the magnetic field and
$\gamma$ is the particle Lorentz factor. More precisely, only the part
of the Lorentz factor related to the particle velocity perpendicular
to the magnetic field should be included, but we ignore this detail
for a rough estimate. In terms of the Larmor radius $r_B$, one can
write $p_\perp=eBr_B/c$, with $e$ the electron charge, to obtain
another form of the adiabatic invariant, $B r_B^2=\const$, i.e.\
magnetic flux through the electron's orbit remains constant. For
$B\propto\rho$, we then obtain $r_B\propto\rho^{-1/2}$ and
$p_\perp\propto\rho^{1/2}$, or $\gamma\propto\rho^{1/2}$.

Thus, compression of magnetic field leads to an increase in the
Lorentz factor of the cosmic-ray electrons, $\gamma\propto\rho^{1/2}$.
If the initial range of the Lorentz factors is
$\gamma_\mathrm{min}\up\leq\gamma\leq\gamma_\mathrm{max}\up$,
compression transforms it into
$\gamma_\mathrm{min}\down\leq\gamma\leq\gamma_\mathrm{max}\down$ such
that
$\gamma_\mathrm{min}\down/\gamma_\mathrm{min}\up=(\rho\down/\rho\up)^{
1/2}$.  Of course, the total number density of cosmic-ray particles
does not change because the cosmic-ray gas is not compressed. Adopting
a power-law spectrum of cosmic-ray electrons,
\[
n_\gamma\, \dd \gamma = K_\gamma \gamma^{-s}\, \dd\gamma\;,
\]
where $n_\gamma\,\dd\gamma$ is the number of relativistic electrons
per unit volume in the range $(\gamma,\gamma+\dd\gamma)$, the total
number density of the particles follows as
$\int_{\gamma_\mathrm{min}}^{\gamma_\mathrm{max}}
K_\gamma\gamma^{-s}\,\dd\gamma\simeq
\gamma_\mathrm{min}^{1-s}K_\gamma=\const$, where we have assumed that
the energy spectrum is broad enough to have
$\gamma_\mathrm{max}\ll\gamma_\mathrm{min}$ and $s>1$. However, the
energy of each cosmic-ray particle increases as $\rho^{1/2}$: the
energy shifts along the energy (or $\gamma$) axis, and \[
K_\gamma\propto\rho^{(s-1)/2}\;, \] i.e., the number density of
particles with given $\gamma$ increases with $\rho$.

Now we can estimate the effect of compression on the synchrotron emissivity
observed at a fixed frequency, $\nu=\const$ and fixed frequency interval
$\dd\nu=\const$. Denoting $\epsilon_I(\nu)$ the arm--interarm contrast in $I(\nu)$ we have
\[
\epsilon_I(\nu)=\frac{I^{(d)}}{I^{(u)}}
\propto\rho^s\;,
\]
since $I\propto K_\gamma B^{(s+1)/2}\propto \gamma_\mathrm{min}^{s-1} B^{(s+1)/2}$.
Note that the Lorentz factor of the electrons which radiate at a
fixed frequency $\nu\simeq\gamma^2 B=\const$ reduces as $B$ increases,
$\gamma(\nu)\propto B^{-1/2}$; this also leads to an increase in the number of
cosmic-ray electrons radiating at a given frequency after compression since 
$n_{\gamma}\propto\gamma^{-s}$ and $s>0$.

With $s\simeq3$ and the arm-interarm density contrast of about four,
the number of cosmic-ray electrons with a given $\gamma$ is
proportional to $K_{\gamma}\propto\rho^{(s-1)/2}\propto\rho$, and the
synchrotron intensity in the spiral arms would then be 50--100 times
stronger than between the arms. There are other reasons to expect
enhanced synchrotron emission in the arms as supernova remnants, sites
of cosmic ray acceleration, are localized in the arms. However,
Table~\ref{con} clearly shows that such an enormous contrast is not
observed.

It is plausible that cosmic rays are rather uniformly distributed in
galactic discs and only weakly perturbed by the spiral arms
\citep[Section 3.10 of][]{Ginzburg:1990}. During their lifetime within
the galaxy, $\tau\simeq3\times10^7\yr$, the cosmic ray particles
become well mixed over distances of order
$(D\tau)^{1/2}\simeq2$--$3\kpc$, where
$D\simeq4\times10^{28}\cm^2\s^{-1}$ is the cosmic ray diffusivity.
This scale exceeds the width of spiral arms, so diffusion can
significantly reduce the arm-interarm contrast, but it can hardly
result in the almost uniform distribution of cosmic rays which would
explain the low contrasts in Table~\ref{tab:arminter}. Anyway, even
assuming that the cosmic ray intensity is the same within the ams and
between them, the compression of magnetic field by a factor of four
would result in an enhancement of the synchrotron emissivity by a
factor 16 for $s=3$, and this is already larger than what is observed. Note that although the contrast in total radio emission given in Table~\ref{tab:arminter} includes the thermal radio emission, this is concentrated in the arms as can be seen in H$\alpha$ images \citep{Greenawalt98} and so the actual contrast in synchrotron emissivity will be lower than the contrast in total radio emissivity.

\subsection{Telescope resolution and the width of the compressed region}
\label{subsec:beam}

One factor that can help to explain the lower than expected
arm-interam contrast in total radio emission (which for simplicity we
assume to be all synchrotron emission) is the possibility that the
compressed region is narrow compared to the width of the spiral arms.
Given the complex processes that take place as gas passes through the
arms, such as the formation of molecular clouds and star formation, it
is plausible that shock compression is followed by a
decompression before the gas leaves the arms. Then, if the compressed
region is narrower than the telescope beam the observed contrast
between the arm and interarm will be reduced. We can estimate the
width of the compressed region that is compatible with the
observations in Table~\ref{con} using
\[
\epsilon_I=\epsilon_I^0\frac{w}{D},
\]
where $\epsilon_I\simeq 5$ is the observed arm-interam contrast in
total radio emission in the inner spiral arms (which we assume to be
all synchrotron emission to make a conservative estimate), 
$\epsilon_I^0\simeq 50$ is the expected
contrast due to compressive effects in an adiabatic shock, $D\simeq
150\pc$ is the beamwidth at our highest resolution and $w$ is the
width of the compressed region. Using the values of the parameters
just quoted we obtain $w\simeq 15\pc$.

\citet{Patrikeev06} showed that the ridge of strongest polarized
emission (tracing the peak of the compressed magnetic field) is
generally shifted upstream of the ridge of strongest CO emission
(tracing the highest neutral gas density) in M51. We can assume that
this shift is due to a spiral-shock triggering the formation of
molecular clouds. The dense clouds will fill only a small fraction of
the volume occupied by the ISM and may become decoupled from the
magnetic field threading the diffuse ISM during their formation, as
originally suggested in the case of M51 by \citet{Tilanus:1988}, by
\citet{Beck:2005} for barred galaxies and with a plausible mechanism
for the separation outlined in \citet{Fletcher:2009}. We expect that
the expansion of the compressed magnetic field will begin once the
clouds have formed. Thus after a time $\tau_c\sim 10^6\yr$ the ridge
in strong radio emission will begin to decay. If we estimate the
magnetic field strength to be $B\simeq 20\uG$ (Sect.~\ref{sec:spx})
then an Alfv\'en wave will propagate over the compressed distance
$w=15\pc$ if the density, of the compressed diffuse gas in which the
clouds are embedded, is $n\simeq 8\cmcube$. This density is not
implausible if the upstream diffuse gas density is $n\simeq 2\cmcube$.

So one possible explanation for the observed low arm-interam contrast
in total radio emission, compared to that  expected from a simple
consideration of cosmic ray energies, is that the compressed region is
narrower than the beam. In this case we estimate that the ridge in
compressed magnetic field, along the upstream edge of the spiral arm,
will be a few tens of parsecs wide. This possibility can be tested
using higher resolution observations. The cosmic rays are not the only
source of the arm-interam contrast in radio emission though: next we
consider the effect of a large-scale shock on the regular and random
components of the magnetic field.

\subsection{Compression of a partially ordered magnetic field}
\label{subsec:comp}

We consider
compression of a partially ordered magnetic field by the spiral
shock. We assume that the random magnetic field upstream of the
shock is statistically isotropic but one-dimensional compression
makes it anisotropic and so it contributes to the polarized radio
emission but not to Faraday rotation. We assume that both the spiral
arms and the field lines of the large-scale magnetic field are
logarithmic spirals with the pitch angles $p\arm$ and $p_b$,
respectively. An acceptable estimate is $p\arm\approx p_b\approx-20^\circ$, where
negative values correspond to a trailing spiral.

The formulae required to calculate the compression of a magnetic
field with both regular and isotropic random components and the
associated total and polarized synchrotron emission and Faraday
rotation are derived in Appendix~\ref{CPOMF}. Using these equations
we now consider two cases that encompass the range of observed
contrasts in polarized emission in M51: the inner spiral arms, where
there is a strong arm--interarm contrast in polarization of at least
$\epsilon_{PI}\sim 4$, and the average contrast of $\epsilon_{PI}\sim 1$
which is more representative of the greater radii.
The upstream ratio of random to regular
magnetic field strength $b/B$ and the increase in gas density
$\epsilon_n=4$, in regular field $\epsilon_B$ and in random field
$\epsilon_b$ fields due to the shock are fixed and then the
consequential increases in nonthermal $\epsilon_I$ and polarized
$\epsilon_{PI}$ emission across the shock are calculated.

Firstly the inner arms. Here we can obtain a reasonable match with
the maximum observed arm--interarm contrasts $\epsilon_I\simeq 5$
(assumed to be mostly synchrotron emission) and $\epsilon_{PI}\simeq 7$
(a lower limit due to weak interarm intensity at $4\arcsec$
resolution). With the parameters $\epsilon_B=1$, i.e. the regular
field is not increased by the shock --- this can be justified, for
example, if the regular magnetic field becomes detached from dense
clouds as they form \citep{Beck:2005} --- and $\epsilon_b=2.7$ we
obtain expected arm--interarm contrasts in synchrotron emission of
$3.3$ and polarized emission of $4.8$. If $\epsilon_B>1$ then the
expected increase in polarization becomes much larger. For
example, for $\epsilon_B=\epsilon_b=2.7$ we obtain
$\epsilon_{PI}=11$: since we only have a lower limit on the observed
$\epsilon_{PI}$ we cannot rule out this possibility. We conclude that
in the innermost spiral arms ($r<1.6\kpc$) anisotropic
random magnetic fields produced by compression of the interstellar
gas in the spiral arms can account for the observed increases in
total and polarized radio emission.

Now for the average arm--interarm contrasts. Here, since the average
contrast in polarization is $\epsilon_{PI}\simeq 1$, this requires no
increase in regular magnetic field in the arm and also
no anisotropy of the random magnetic field, but simultaneously
we require a small increase in synchrotron emission
$\epsilon_I=2.2$. This is not possible in our model; it is also
unlikely to happen when galaxies contain strong spiral density
waves. The closest we can come is to set $\epsilon_B=1$ coupled with
a weak upstream random field $b^2/B^2=0.2$ and strong compression of
the random field $\epsilon_b=4.5$ to produce the required contrast
in total emission. Then we can obtain arm--interarm contrasts of
$2.3$ in total synchrotron emission and $1.7$ in polarization.
However, $b^2/B^2=0.2$ is strongly contradicted by the equipartition
estimates of the various field strengths in Sect.~\ref{sec:spx},
which likely overestimate the strength of the regular field as
discussed in Sect.~\ref{sec:dp} and \ref{sec:B}.

To summarise this subsection: enhancement of the regular and random
magnetic field components parallel to a large-scale spiral shock can
\emph{partly} account for the observed arm-interam contrasts in radio
emission but no single set of parameters is compatible with the full
range of the observations.

\subsection{Decompression of an isotropic magnetic field}
\label{subsec:decomp}

Finally, we consider the \emph{decompression} of an isotropic field as
the magnetized gas flows from a high density to a low density region.
This decompression also leads to the generation of an anisotropic
random magnetic field, as the field component parallel to lines of
constant density increases while the perpendicular component is
unaffected, similar to the compressive case. But decompression will
also help to lower the arm-interam contrast in nonthermal emission,
thus alleviating one of the difficulties encountered in the previous
subsection, if the random field in the arms is (partly) isotropized:
this can readily occur due to turbulence being driven by star
formation activity and the expansion of supernova remnants.

First consider the case of straightforward compression. Let us define
the $x$-direction as perpendicular to the shock and the $y$-direction
as parallel: so only $y$-components of the magnetic field are affected
by the shock. Thus
\[
\bra{b_x^{2(d)}}=\bra{b_x^{2(u)}}=\frac{1}{3}b^{2(u)},
\]
and
\[
\bra{b_y^{2(d)}}=\epsilon_{\rho}^2\bra{b_y^{2(u)}}=\frac{1}{3}\epsilon_{\rho}^2 b^{2(u)},
\]
where $(u)$ and $(d)$ refer to upstream and downstream and $\vec{b}$
is the total random field. Now the plane-of-sky component of the
random field is $b_{\perp}^2=b_x^2+b_y^2$ so we obtain
\[
\frac{b_{\perp}^2(d)}{b_{\perp}^2(u)}=\frac{1}{2}(1+\epsilon_{\rho}^2)\simeq 8,
\]
for an adiabatic shock, where $\epsilon_{\rho}=4$. So we expect
adiabatic compression in a simple, plane parallel shock to produce a
contrast in nonthermal emission $\epsilon_I\simeq 8$, in the case that
the cosmic rays are smoothly distributed. This is a simplified version
of part of the calculation described above and in detail in
Appendix~\ref{CPOMF}.

Now consider the case of decompression. We follow a similar calculation but here
\[
\bra{b_y^{2(d)}}=\frac{1}{\epsilon_{\rho}^2}\bra{b_y^{2(u)}}=\frac{1}{3\epsilon_{\rho}^2} b^{2(u)},
\]
which leads to
\[
\frac{b_{\perp}^2(d)}{b_{\perp}^2(u)}=\frac{1}{2}(1+\frac{1}{\epsilon_{\rho}^2})\simeq \frac{1}{2}.
\]
In this case the arm is the upstream region so the expected
arm-interam contrast in nonthermal emission is $\epsilon_I\simeq 2$
which is close to the average contrast observed (Table~\ref{con}).

\subsection{Summary: compression and arm-interarm contrasts}

In Sect.~\ref{sec:dp} and \ref{sec:B} we have shown that much of the
polarized emission detected across the disc of M51 must come from
anisotropic random fields. Combined with the problem described in
Sect.~\ref{subsec:comp} above, of how isotropic random fields can be
compressed in a spiral shock but not produce an increase in polarized
emission, we are led to the view that anisotropic random fields are
already present in interarm regions, perhaps as a result of enhanced
localised shear or decompression. \citet{Patrikeev06} showed that the
orientation of the magnetic field in M51 varies in azimuth by $\pm
15\degr$ and in the interarm generally has a different pitch angle to
the CO spiral arm, only becoming well aligned with the CO arm at its
location. In this case compression of the already anisotropic field in
the spiral shock will only weakly amplify the random field and hence
lead to a weak change in polarized emission.

We conclude that the roughly constant average polarized emission
across the arms and interam region cannot be easily explained with
simple models of shock compression of magnetic field, if one
simultaneously considers the weak contrasts that are observed in
the total emission. The arm-interarm contrast in the radio
emission probably results from a complex interplay of compression
\emph{and} decompression of the dominant random field, occurring as
the ISM undergoes phase changes on its passage through the arms. In
addition the thickness of the compressed regions compared to the width
of the beam likely plays a role.

\section{Summary and Conclusions}

\begin{itemize}
\item Polarized emission ($PI$) is present throughout most of
the disc of M51. In some regions the strongest $PI$ coincides with
the location of the strong spiral arms as seen in CO emission. In
other regions $PI$ is concentrated in the interarm region, forming
structures up to $\sim 5\kpc$ in size, reminiscent of the
$\sim10\kpc$-scale magnetic arms observed in NGC~6946
\citep{Beck96}. The origin of these magnetic arms is still unknown.

\item The observed polarization angles trace spiral patterns with
pitch angles similar to, but not always the same as, the gaseous
spiral arms. The apparently ordered large-scale magnetic field
responsible for the well aligned polarization angles does not
produce a systematic pattern in Faraday rotation, leading us to
conclude that a large fraction of the polarized emission is caused
by anisotropic small-scale magnetic fields (where small-scale refers
to sizes smaller than the beam, typically $300$--$600\pc$):
anisotropic random fields, whose anisotropy is caused by a large
scale process \hl{(for example, compression and/or shear)} and so is aligned over large distances, can produce
well ordered polarization angles but a random Faraday rotation
distribution.

\item Faraday depolarization, caused by Faraday dispersion due
to turbulent magnetic fields, leads to strong depolarization of
the \wav{20} polarized emission from the disc. From the observed
fluctuations in Faraday depolarization we were able to estimate a
typical \hl{diameter} of a turbulent cell
as $\sim 50\pc$.

\item Fourier filtering followed by averaging in sectors is necessary
to reveal the contribution of the regular (or mean) magnetic field
to the observed polarization signal. This allowed us to fit a model
of the 3D regular magnetic field of M51 to the observations of
polarization angles at \wwwwav{3}{6}{18}{20}. Due to the strong
depolarization at \wwav{18}{20} we were able to identify two
different regular magnetic field patterns. In the thermal disc the regular
field can be described as a combination of $m=0+2$ azimuthal modes,
with the $m=0$ mode being the strongest: this combination can
be the result of the strong two-armed spiral pattern modifying a
dynamo generated $m=0$ mode (the easiest to excite according to
mean-field galactic dynamo theory). The pitch angle of the $m=0$
mode is similar at all radii. In the halo the observed
polarization angles at \wwav{18}{20}, whose emission from the thermal disc is
heavily depolarized, reveal a $m=1$ azimuthal geometry for the
regular magnetic field. The origin of this halo field is unclear.

\item The arm--interarm contrast in gas density and radio emission was
compared to a model where a regular and (isotropic) random magnetic
field is compressed by shocks along the spiral arms. We found that
where a strong arm--interarm contrast in $PI$ is observed, in the inner
arms $r\lesssim 1.6\kpc$, amplification of the random magnetic field by
compression can successfully explain the data, provided that the
regular magnetic field is not significantly increased. This constraint
is similar to that obtained for two barred galaxies in
\citet{Beck:2005}, where it was proposed that the regular magnetic
field de-couples from molecular clouds as they rotate and collapse. We
were unable to explain the average arm--interarm contrast in total and
polarized radio emission, typical for much of the galaxy at $r>2\kpc$,
by a model involving shock compression of magnetic fields. Even when
the regular magnetic field is not compressed some increase in $PI$ in
the arms is expected from compression of the random field, whereas the
average arm--interarm contrast in $PI$ is about one: this problem could
be resolved if the random field is isotropic in the arms and becomes
anisotropic due to decompression as it enters the interam.
Alternatively, the compressed region of magnetic field may be
sufficiently narrow (with a width of about $20\pc$), as might occur if
the molecular clouds are de-coupled from the synchrotron emitting gas,
to reduce the arm--interarm contrast by the required degree.

\end{itemize}

%
%
\section*{Acknowledgments}

We thank W. Reich for suggesting the comparison
of RM structure functions of our maps and a Milky Way model and for kindly
making the calculations. We thank K. Ferri\`ere for helpful
discussions and M. Krause for a careful reading of the manuscript. 
A.F.\ and A.S.\ gratefully acknowledge financial support
under the Leverhulme Trust grant F/00 125/N and the STFC grant
ST/F003080/1. This research made use of NASA's Extragalactic database
(NED) and Astrophysical Data System (ADS).

%
%
\appendix
\section{Parameters of the fitted regular magnetic fields}
\label{app:Bmodel}

In Table~\ref{tab:fit} we give the parameters of the fitted regular magnetic
field models discussed in Section~\ref{sec:B}. Although a component of the
regular field perpendicular to the disc plane ($R_z$ in Eq.~(\ref{eq:Bmod}))
is allowed in the model and we searched for fits using this component, a
vertical field was not required to obtain a good fit in either ring. The
greater of the standard deviation and the noise within a sector was taken as
the error in polarization angle. 

The regular magnetic field is modelled as
\begin{eqnarray}
  R_r & = &  R_0\sin p_0 + R_1\sin p_1\cos(\phi-\beta_1)\nonumber \\
      & & +\, R_2\sin p_2\cos(2\phi-\beta_2),\nonumber \\
  R_{\phi} & = & R_0\cos p_0 + R_1\cos p_1\cos(\phi-\beta_1) \nonumber\\
      & & +\, R_2\cos p_2\cos(2\phi-\beta_2),\label{eq:Bmod}\\
  R_z & = & R_{z0} + R_{z1}\cos(\phi-\beta_{z1}) + R_{z2} \cos(2\phi-\beta_{z2}),\nonumber\\
  R_{hr} & = & R_{h0}\sin p_{h0} + R_{h1}\sin p_{h1}\cos(\phi-\beta_{h1}),\nonumber \\
  R_{h\phi} & = & R_{h0}\cos p_{h0} + R_{h1}\cos p_{h1}\cos(\phi-\beta_{h1}),\nonumber
 \end{eqnarray}
where $R_i$ is the amplitude of the $i$'th mode in units of $\FRM$,
$p_i$ is its pitch angle, $\beta_i$ ($i\ge1$) determines the
azimuth where the corresponding non-axisymmetric mode is maximum and the subscript $_{h}$ denotes the components of the halo field.
The magnetic field in each mode of this model is approximated by a
logarithmic spiral, $p_i=\mbox{const}$, within a given ring.
However, the superposition of such modes with different pitch angles
leads to deviations from a logarithmic spiral. For further details
of the method, see \citet{Berkhuijsen:1997} and
\citet{Fletcher:2004}.

The foreground Faraday rotation due to the magnetic field of the Milky
Way $R_\mathrm{fg}$ was also included in the fitting; we expect
$R_\mathrm{fg}$ to be the same in all rings and this provides a useful
consistency check to the results of independent fits to the four
rings. A logarithmic spiral pattern has the same pitch angle in all
rings. $S$ is the residual of the fit \hl{and} $S_\lambda$ \hl{the residual at a given wavelength.} The appropriate $\chi^2$ value,
at the $95$\% confidence level, is shown for the number of fit
parameters and data points. A fit is statistically acceptable if
$S\leq\chi^2$ and the Fisher criteria, that tests if $S$ is unduly
influenced by a good fit to a single wavelength, is satisfied. The
$\chi^2$ values vary from ring to ring (even when the same number of
fit parameters are used) as some sectors are excluded from the model,
either because the noise in the sector exceeds the standard deviation
of the measurements or because the sector represents an outlier from
the global pattern. 

\begin{figure}
\centering
\includegraphics[width=0.49\textwidth]{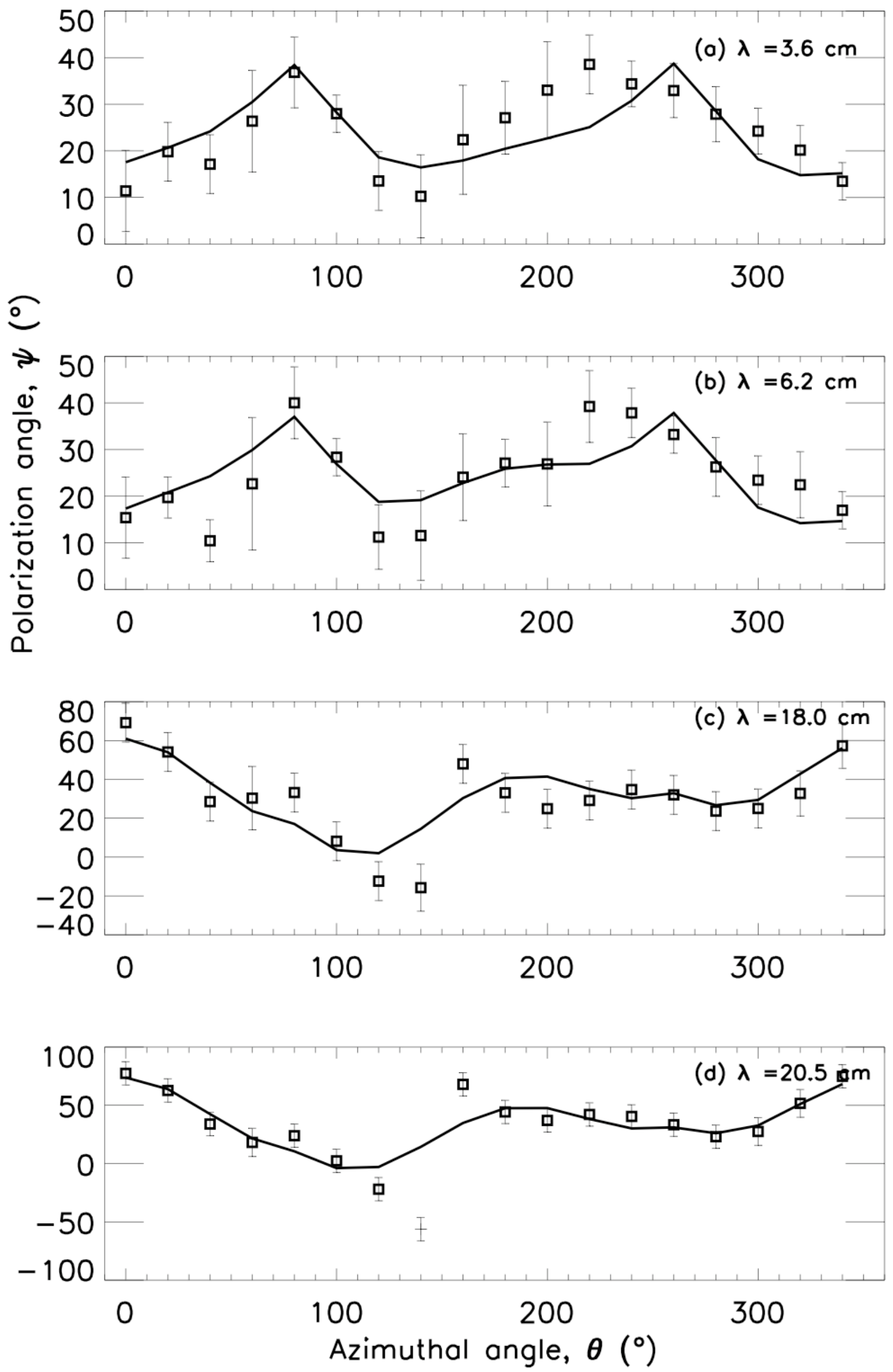}
\caption{Polarization angles
    ($\psi$, measured from the local radial direction in the plane of
    M\,51) against azimuth in the galaxy plane ($\theta$) for the ring
    2.4--3.6$\kpc$. Fit (solid line) and observations (squares with
    error bars, horizontal lines with error bars show data points excluded
    from the fit) are shown for \wav{3}, \wav{6}, \wav{18} and
    \wav{20} from top to bottom. The error bars show the $1\sigma$
    deviations.}
\label{fig:ring1}
\end{figure}

\begin{figure}
\centering
\includegraphics[width=0.49\textwidth]{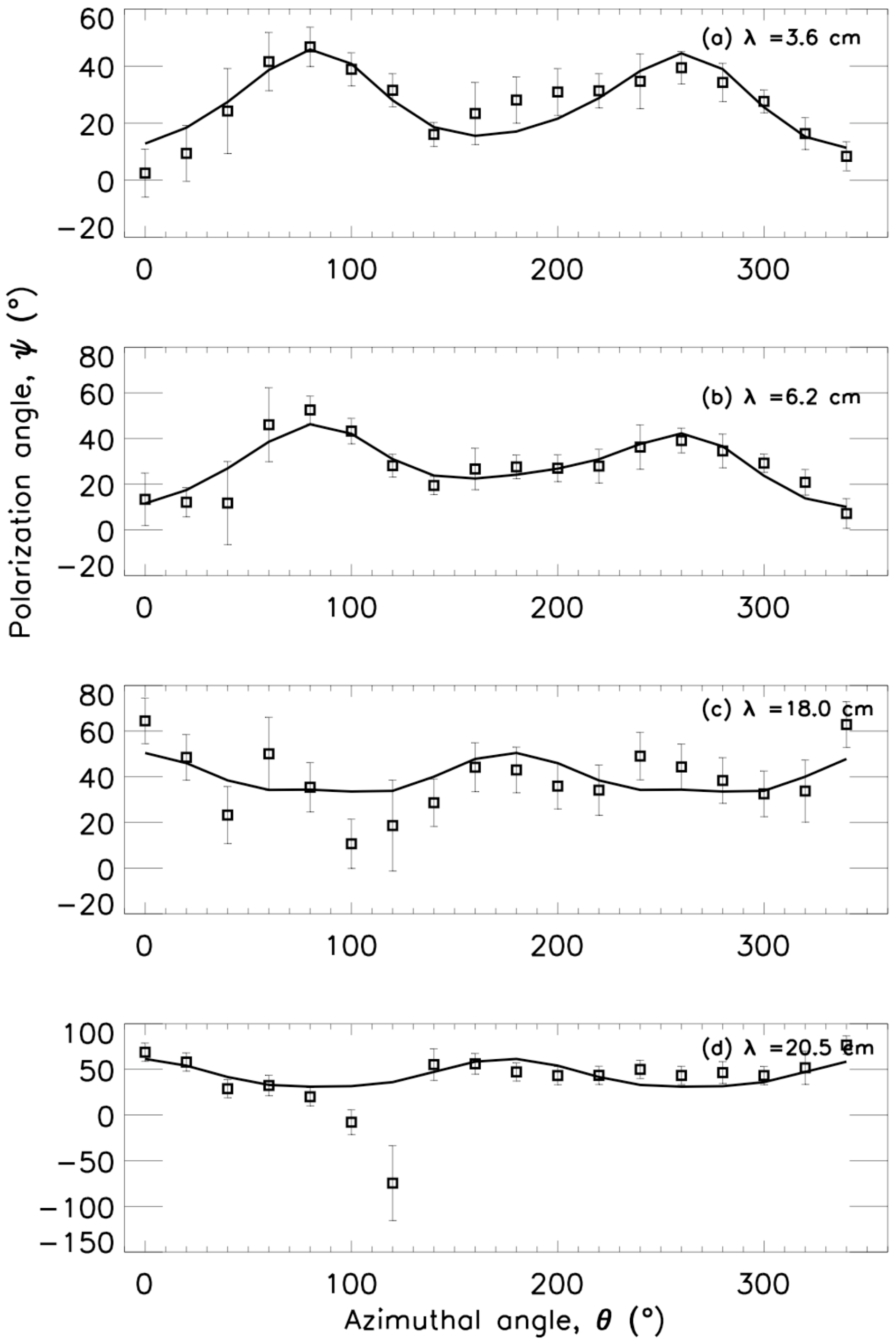}
\caption{As in Fig.~\ref{fig:ring1} but for the ring 3.6--4.8$\kpc$.}
\label{fig:ring2}
\end{figure}

\begin{figure}
\centering
\includegraphics[width=0.49\textwidth]{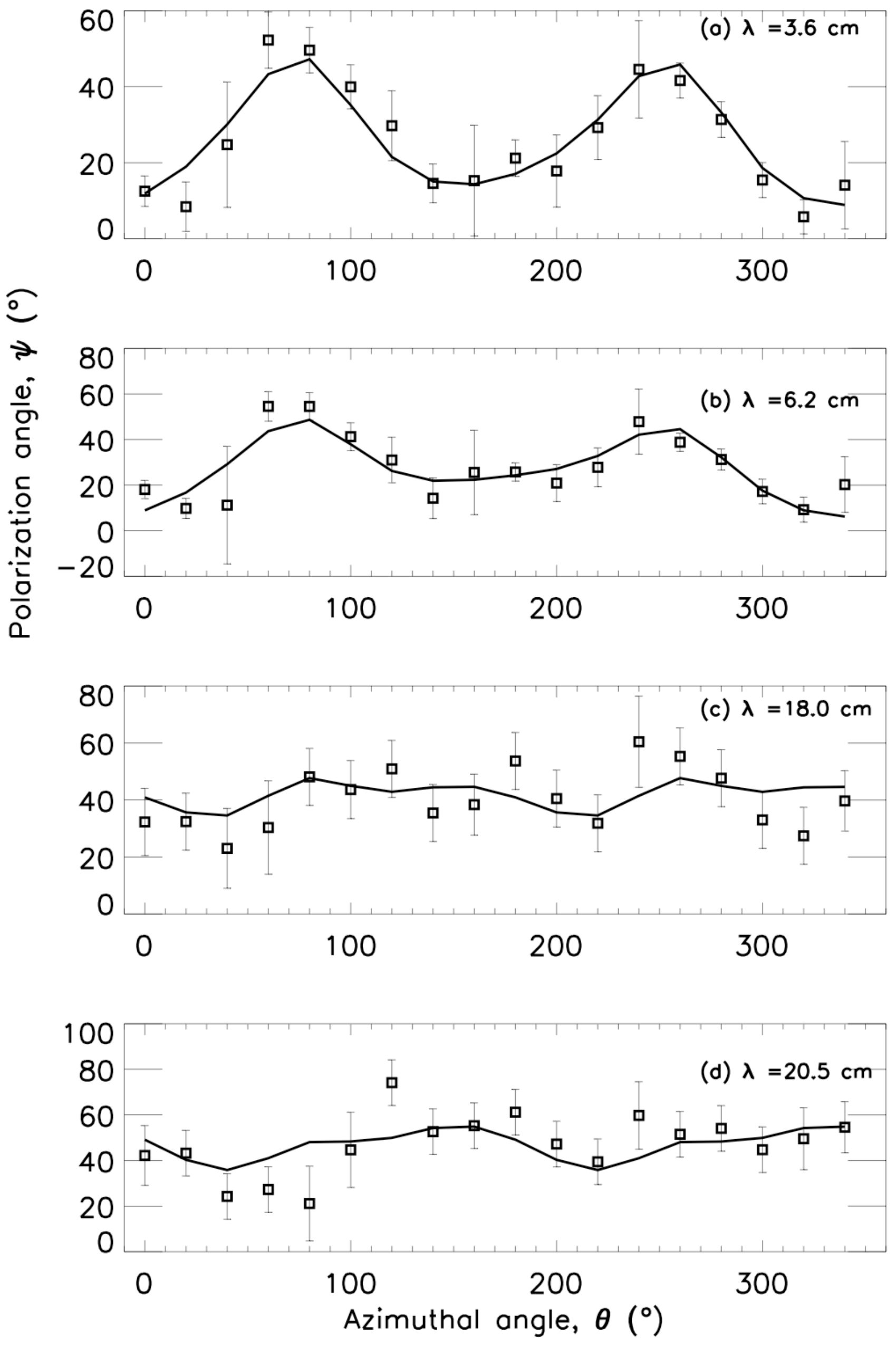}
\caption{As in Fig.~\ref{fig:ring1} but for the ring 4.8--6.0$\kpc$.}
\label{fig:ring3}
\end{figure}

\begin{figure}
\centering
\includegraphics[width=0.49\textwidth]{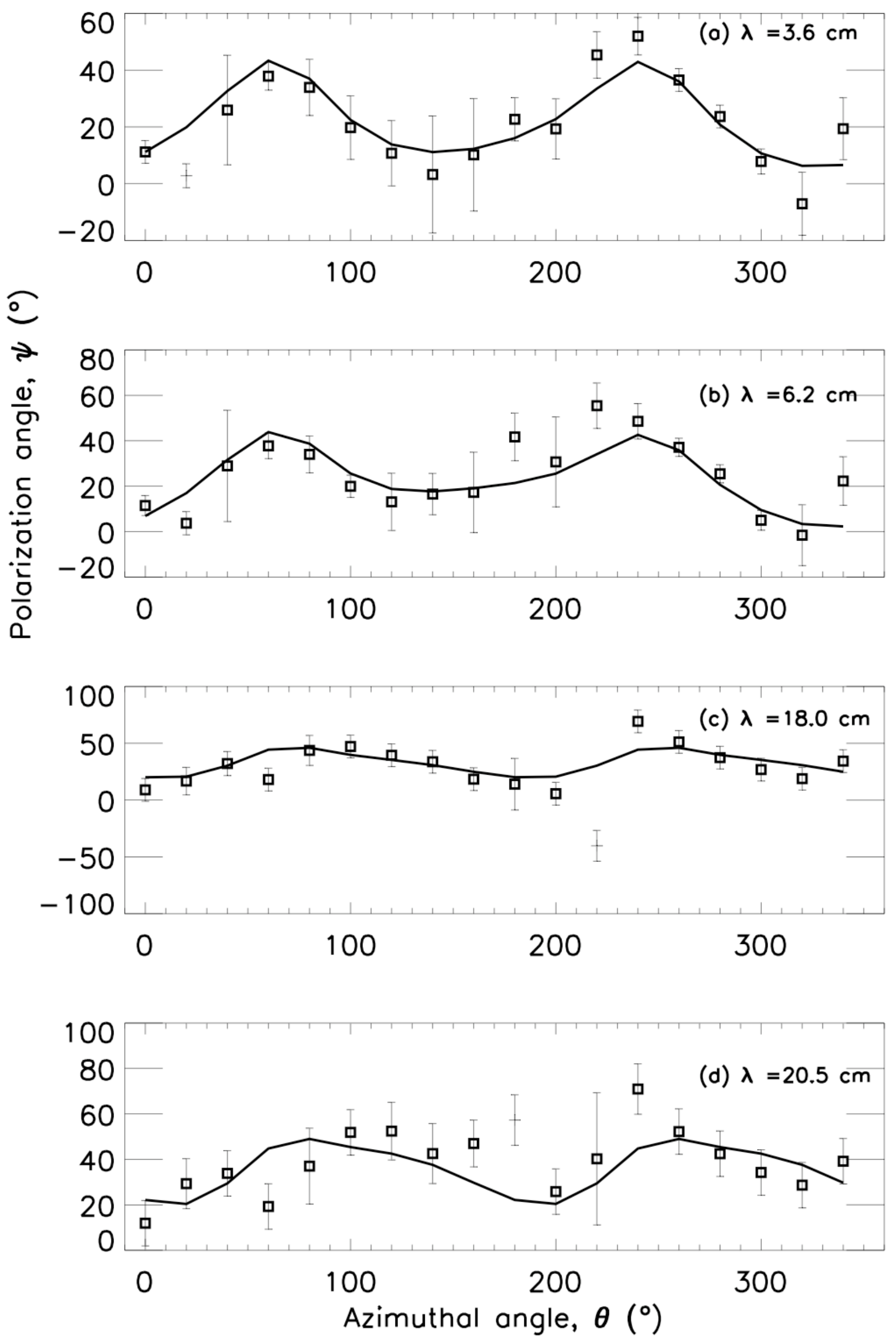}
\caption{As in Fig.~\ref{fig:ring1} but for the ring 6.0--7.2$\kpc$.}
\label{fig:ring4}
\end{figure}

In Figs.~\ref{fig:ring1}, \ref{fig:ring2}, \ref{fig:ring3} and
\ref{fig:ring4} we show the observed sector-averaged polarization
angles and the fitted model for each ring. The fit quality is
excellent at \wav{3} and \wav{6} for all rings, but in the inner two
rings sharp discontinuities in the \wav{20} polarization angles around
$\theta\simeq 120\degr$ cannot be accommodated by the model. Since the
parameters of the fitted halo field are largely determined by the data
at the longer wavelengths we are therefore less satisfied with the
parameters of the fitted halo field in these rings than with those of
the disc field, as we discuss in Sect.~\ref{sec:B}.

\begin{table*}
\caption{Parameters of fitted model for M51, with
notation as in Eq.~\ref{eq:Bmod}. The index $h$ refers to the ``halo''
field. The residuals at the wavelengths \wwwwav{3}{6}{18}{20} are given in brackets.}
\label{tab:fit} 
\begin{tabular}{llllll}
\hline
  & & $2.4<r<3.6\kpc$ & $3.6<r<4.8\kpc$ & $4.8<r<6.0\kpc$ & $6.0<r<7.2\kpc$ \\
\hline
$R_\mathrm{fg}$ & $\FRM$ &
  $4$\,\scriptsize{$\pm2$} & $5$\,\scriptsize{$\pm4$} &
  $2$\,\scriptsize{$\pm5$} & $4$\,\scriptsize{$\pm1$} \\
\\
$R_0$ & $\FRM$ &
  $-46$\,\scriptsize{$\pm3$} & $-57$\,\scriptsize{$^{+11}_{-20}$} &
  $-76$\,\scriptsize{$^{+14}_{-27}$} & $-76$\,\scriptsize{$\pm2$} \\
$p_0$ & deg &
  $-20$\,\scriptsize{$\pm1$} & $-24$\,\scriptsize{$\pm4$}  &
  $-22$\,\scriptsize{$\pm4$} & $-18$\,\scriptsize{$\pm1$}  \\
$R_2$ &  $\FRM$ &
  $-33$\,\scriptsize{$\pm2$} & $-25$\,\scriptsize{$^{+7}_{-5}$} &
  $-40$\,\scriptsize{$^{+11}_{-9}$} & $-44$\,\scriptsize{$\pm2$} \\
$p_2$ & deg &
  $-12$\,\scriptsize{$\pm2$} & $16$\,\scriptsize{$^{+23}_{-15}$} &
  $8$\,\scriptsize{$^{+18}_{-12}$} & $3$\,\scriptsize{$\pm2$} \\
$\beta_2$ & deg &
  $-8$\,\scriptsize{$\pm5$} & $-6$\,\scriptsize{$\pm9$} &
  $-14$\,\scriptsize{$\pm10$} & $-25$\,\scriptsize{$\pm2$} \\
\\
$R_{h0}$ &  $\FRM$ &
  $23$\,\scriptsize{$\pm6$} & & & \\
$p_{h0}$ & deg &
  $-43$\,\scriptsize{$\pm13$} & & &  \\
$R_{h1}$ &  $\FRM$ &
  $76$\,\scriptsize{$\pm11$} & $77$\,\scriptsize{$\pm26$} &
  $57$\,\scriptsize{$\pm26$} & $43$\,\scriptsize{$\pm14$}  \\
$p_{h1}$ & deg &
  $-45$\,\scriptsize{$\pm5$} & $-49$\,\scriptsize{$\pm12$} &
  $-50$\,\scriptsize{$^{+29}_{-23}$} & $-90$\,\scriptsize{$\pm3$} \\
$\beta_{h1}$ & deg &
  $44$\,\scriptsize{$\pm5$} & $30$\,\scriptsize{$\pm14$} &
  $-3$\,\scriptsize{$^{+23}_{-29}$} & $-16$\,\scriptsize{$\pm3$} \\
\hline
$S_{\lambda}$ & (\wwwwav{3}{6}{18}{20}) & $(14, 21, 21, 21)$ & $(9, 9, 20, 28)$ & 
  $(10, 18,  11, 16)$ & $(10, 26, 21, 21)$\\
$S$&& $77$ & $66$ & $55$ & $78$ \\
$\chi^2$ && $79$ & $83$ & $83$ & $79$ \\
\hline
\end{tabular}\vfill
\vspace{0.2cm}
One data point excluded in the ring $2.4<r<3.6\kpc$ at: \wav{20}, $\phi=140\degr$. \\
Three data points excluded in the ring $6.0<r<7.2\kpc$ at: \wav{3}, $\phi=20\degr$; \wav{18}, $\phi=220\degr$; \wav{20}, $\phi=180\degr$.
\end{table*}


\section{Compression of a partially ordered magnetic field in a spiral
arm}
\label{CPOMF}

We introduce a Cartesian frame in the sky
plane $(\tilde x,\tilde y,\tilde z)$ centered at the galaxy center
with the $\tilde x$-axis directed towards the western end of the
major axis and the $\tilde y$-axis, in the northern direction; the
$\tilde z$-axis is the directed towards the observer (and in the
general direction of the galaxy's north pole). We also introduce
galaxy's Cartesian frame $(x,y,z)$ where the $x$- and $\tilde
x$-axes coincide and the $z$-axis is also directed towards the
galaxy's north pole. Magnetic field components in the two frames are
related by \citep{Berkhuijsen:1997}:
\begin{eqnarray} B_{\tilde x}&=&B_x,\nonumber\\
B_{\tilde y}&=&B_y\cos i+B_z\sin i,\label{trans}\\
B_{\tilde z}&=&-B_y\sin i+B_z\cos i,\nonumber
\end{eqnarray}
where $i$ is the galaxy's inclination angle, and we include $B_z$
for the future convenience (we recall that we can adopt $B_z=0$ in
M51). The galaxy's cylindrical frame $r,\phi,z$ then has the
azimuthal angle measured counterclockwise (along the galaxy's
rotation) from the $x$-axis. And finally, we introduce a local
Cartesian frame of the spiral shock $(x',y',z')$, with the
$x'$-axis perpendicular to the shock and directed from the interarm
region into the spiral arm, the $y'$-axis parallel to the shock, so
that the $z'$-axis complements them to a right-handed triad i.e., is
directed towards the north pole of the galaxy. Then the angle
between the $x'$- and $x$-axes is
\begin{equation}\label{theta}
\theta=\phi-p\arm.
\end{equation}

It is convenient to specify the upstream large-scale magnetic field in the
galaxy's cylindrical frame $\vec{\Br}=B(\sin p_B,\cos p_B,0)$, where we
neglect the vertical field component (see Sect.~\ref{sec:B}). Then the unit
normal to the shock, in the galaxy's frame is given by
$\widehat{\vec{n}}=\widehat{\vec{x}'}=(\cos\theta,\sin\theta,0)$, the tangent
vector is given by
$\widehat{\vec{t}}=\widehat{\vec{y}'}=(-\sin\theta,\cos\theta,0)$,  and the
regular magnetic field components normal and tangent to the shock then follow,
respectively, as
\[
B_{x'}=B\sin(p_b-p\arm),\quad B_{y'}=B\cos(p_b-p\arm).
\]

It is now easy to see that the compressed large-scale magnetic field in the
spiral arm has the components
\[
B\down_{x'}=B\up_{x'},\quad B\down_{y'}=\epsilon_n B\up_{y'},
\]
where $\epsilon_n=n\down/n\up$ is the gas density compression ratio.
Now we can transform both the original and compressed fields to the galaxy's
frame by rotating it by the angle $-\theta$,
$B_{x}=B_{x'}\cos\theta-B_{y'}\sin\theta,$
$B_{y}=B_{x'}\sin\theta+B_{y'}\cos\theta,$
$B_{z}=B_{z'},$  and then to the sky frame using
Eq.~(\ref{trans}), where the $\tilde x$ and $\tilde y$ component contribute to
the magnetic field in the plane of the sky and $B_{\tilde z}$ is directed
along the line of sight:
\[
\vec{B}_\perp=(B_{\tilde x}, B_{\tilde y}),\quad B_\parallel =B_{\tilde z}.
\]

Similar relations can be written for the random magnetic field
$\vec{\bt}$, but now we cannot neglect the $z$-component of the
random magnetic field in the galaxy frame. The compressed random
field in the shock frame is given by
\[
\vec{\bt}\down=(b\up_{x'}, \epsilon_n b\up_{y'},\epsilon_n b\up_{z'})
\]
which can be transformed to the galaxy's frame and then to the sky frame to
obtain
\begin{eqnarray*}
\bt_{\tilde x}&=&b_{x}=b_{x'}\cos\theta-\epsilon_nb_{y'}\sin\theta,\\
\bt_{\tilde y}&=&b_{x'}\sin\theta\cos i+\epsilon_n b_{y'}\cos\theta\cos i
                +\epsilon_n b_{z'}\sin i,\\
\bt_{\tilde z}&=&-b_{x'}\sin\theta\sin i-\epsilon_n b_{y'}\cos\theta\sin i
                +\epsilon_n b_{z'}\cos i.
\end{eqnarray*}
Averaging using these equations then yields
\begin{eqnarray*}
\bra{{\bt_{\tilde x}^2}}\down&=&\frac13{\sigma_b^2}\up
        \left[1+(\epsilon_n^2-1)\sin^2\theta\right],\\
\bra{\bt_{\tilde y}^2}\down&=&\frac13{\sigma_b^2}\up
        \left[1+(\epsilon_n^2-1)(-1\sin^2\theta\cos^2 i)\right],\\
\bra{\bt_{\tilde z}^2}\down&=&\frac13{\sigma_b^2}\up
        \left[1+(\epsilon_n^2-1)(-1\sin^2\theta\sin^2 i)\right],
\end{eqnarray*}
where
$\bra{{\bt_{x'}^2}}\up=\bra{{\bt_{y'}^2}}\up=\bra{{\bt_{z'}^2}}\up=\frac13{\sigma_b^2}\up$
(isotropy of the upstream random magnetic field), and
$\bra{\bt_{x'}\bt_{y'}}\up=\bra{\bt_{x'}\bt_{z'}}\up=\bra{\bt_{y'}\bt_{z'}}\up=0$
(statistical independence of the upstream field components).

Now the arm--interarm contrasts in various observables can be
estimated as follows. For the total synchrotron intensity,
\[
\epsilon_I\simeq
\frac{{B_\perp^2}\down+\bra{\bt_\perp^2}\down}{{B_\perp^2}\up+\bra{\bt_\perp^2}\up}
\,\frac{L_I\down}{L_I\up}\frac{n\down_{\gamma}}{n\up_{\gamma}},
\]
where $L_I$ is the pathlength through the synchrotron layer and
$n_{\gamma}$ the cosmic ray number density. The
polarized emissivity in a partially ordered, anisotropic random
magnetic field and uniform cosmic ray distribution can be calculated
using Eq.~(20) of \citet{Sokoloff98} as
\[
PI\propto(B_{\tilde x}^2-B_{\tilde y}^2
        +\bra{b_{\tilde x}^2}-\bra{b_{\tilde y}^2})^2
        +4B_{\tilde x}^2B_{\tilde y}^2.
\]
Applying this formula to the downstream magnetic field, we obtain
the contrast in polarized intensity $\epsilon_{PI}$ by dividing it by
$PI\up\propto{B_\perp^2}\up$, and also multiplying by the ratio of
pathlengths $L_I\down/L_I\up$ and cosmic ray densities $n\down_{\gamma}/n\up_{\gamma}$.

The Faraday rotation measure can be calculated as $\RM\simeq 0.81\n
B_{\tilde z} L_{\RM}$, where $L_{\RM}$ is an appropriate pathlength,
and its standard deviation is obtained using Eq.~(\ref{eq:sigRM}), and
then the arm--interarm contrast is calculated straightforwardly.


\bsp

\label{lastpage}

\end{document}